\documentclass[final]{IEEEtran}
\usepackage{amsthm,amssymb,graphicx,multirow,amsmath,color,amsfonts}
\usepackage[update,prepend]{epstopdf}
\usepackage[noadjust]{cite}
\usepackage[latin1]{inputenc}
\usepackage{tikz}
\usepackage{bbm} 
\usepackage[nolist]{acronym}
\usepackage{pdfpages}
\usepackage{multirow}
\usepackage{subfig}
\usepackage{comment} 
\def\nb0{{\mathbf{0}}}
\def\nb1{{\mathbf{1}}}

\newtheorem{lemma}{Lemma}

\newtheorem{theorem}{Theorem}
\newtheorem{prop}{Proposition}
\newtheorem{cor}{Corollary}

\newtheorem{approximation}{Approximation}	

\allowdisplaybreaks 

\begin{acronym}

\acro{5G-NR}{5G New Radio}
\acro{3GPP}{3rd Generation Partnership Project}
\acro{AC}{address coding}
\acro{ACF}{autocorrelation function}
\acro{ACR}{autocorrelation receiver}
\acro{ADC}{analog-to-digital converter}
\acrodef{aic}[AIC]{Analog-to-Information Converter}     
\acro{AIC}[AIC]{Akaike information criterion}
\acro{aric}[ARIC]{asymmetric restricted isometry constant}
\acro{arip}[ARIP]{asymmetric restricted isometry property}
\acro{AoA}{angle of arrival}
\acro{AoD}{angle of departure}
\acro{ARQ}{automatic repeat request}
\acro{AUB}{asymptotic union bound}
\acrodef{awgn}[AWGN]{Additive White Gaussian Noise}     
\acro{AWGN}{additive white Gaussian noise}

\acro{APSK}[PSK]{asymmetric PSK} 

\acro{waric}[AWRICs]{asymmetric weak restricted isometry constants}
\acro{warip}[AWRIP]{asymmetric weak restricted isometry property}
\acro{BCH}{Bose, Chaudhuri, and Hocquenghem}        
\acro{BCHC}[BCHSC]{BCH based source coding}
\acro{BEP}{bit error probability}
\acro{BFC}{block fading channel}
\acro{BG}[BG]{Bernoulli-Gaussian}
\acro{BGG}{Bernoulli-Generalized Gaussian}
\acro{BPAM}{binary pulse amplitude modulation}
\acro{BPDN}{Basis Pursuit Denoising}
\acro{BPPM}{binary pulse position modulation}
\acro{BPSK}{binary phase shift keying}
\acro{BPZF}{bandpass zonal filter}
\acro{BSC}{binary symmetric channels}              
\acro{BU}[BU]{Bernoulli-uniform}
\acro{BER}{bit error rate}
\acro{BS}{base station}

\acro{CP}{Cyclic Prefix}
\acrodef{cdf}[CDF]{cumulative distribution function}   
\acro{CDF}{cumulative distribution function}
\acrodef{c.d.f.}[CDF]{cumulative distribution function}
\acro{CCDF}{complementary cumulative distribution function}
\acrodef{ccdf}[CCDF]{complementary CDF}               
\acrodef{c.c.d.f.}[CCDF]{complementary cumulative distribution function}
\acro{CD}{compliance distance}

\acro{CDMA}{Code Division Multiple Access}
\acro{ch.f.}{characteristic function}
\acro{CIR}{channel impulse response}
\acro{cosamp}[CoSaMP]{compressive sampling matching pursuit}
\acro{CR}{cognitive radio}
\acro{cs}[CS]{compressed sensing}                   
\acrodef{cscapital}[CS]{Compressed sensing}
\acrodef{CS}[CS]{compressed sensing}
\acro{CSI}{channel state information}

\acro{CCSDS}{consultative committee for space data systems}
\acro{CC}{convolutional coding}

\acro{DAA}{detect and avoid}
\acro{DAB}{digital audio broadcasting}
\acro{DCT}{discrete cosine transform}
\acro{dft}[DFT]{discrete Fourier transform}
\acro{DR}{distortion-rate}
\acro{DS}{direct sequence}
\acro{DS-SS}{direct-sequence spread-spectrum}
\acro{DTR}{differential transmitted-reference}
\acro{DVB-H}{digital video broadcasting\,--\,handheld}
\acro{DVB-T}{digital video broadcasting\,--\,terrestrial}
\acro{DL}{downlink}
\acro{DSSS}{Direct Sequence Spread Spectrum}
\acro{DFT-s-OFDM}{Discrete Fourier Transform-spread-Orthogonal Frequency Division Multiplexing}
\acro{DAS}{distributed antenna system}
\acro{DNA}{Deoxyribonucleic Acid}

\acro{EC}{European Commission}
\acro{EED}[EED]{exact eigenvalues distribution}
\acro{EIRP}{Equivalent Isotropically Radiated Power}
\acro{ELP}{equivalent low-pass}
\acro{eMBB}{Enhanced Mobile Broadband}
\acro{EMFE}{electromagnetic filed exposure}
\acro{EU}{European union}

\acro{FC}[FC]{fusion center}
\acro{FCC}{Federal Communications Commission}
\acro{FEC}{forward error correction}
\acro{FFT}{fast Fourier transform}
\acro{FH}{frequency-hopping}
\acro{FH-SS}{frequency-hopping spread-spectrum}
\acrodef{FS}{Frame synchronization}
\acro{FSsmall}[FS]{frame synchronization}  
\acro{FDMA}{Frequency Division Multiple Access}

\acro{GA}{Gaussian approximation}
\acro{GF}{Galois field }
\acro{GG}{Generalized-Gaussian}
\acro{GIC}[GIC]{generalized information criterion}
\acro{GLRT}{generalized likelihood ratio test}
\acro{GPS}{Global Positioning System}
\acro{GMSK}{Gaussian minimum shift keying}
\acro{GSMA}{Global System for Mobile communications Association}

\acro{HAP}{high altitude platform}

\acro{IDR}{information distortion-rate}
\acro{IFFT}{inverse fast Fourier transform}
\acro{iht}[IHT]{iterative hard thresholding}
\acro{i.i.d.}{independent and identically distributed}
\acro{IoT}{Internet of Things}                      
\acro{IR}{impulse radio}
\acro{lric}[LRIC]{lower restricted isometry constant}
\acro{lrict}[LRICt]{lower restricted isometry constant threshold}
\acro{ISI}{intersymbol interference}
\acro{ITU}{International Telecommunication Union}
\acro{ICNIRP}{International Commission on Non-Ionizing Radiation Protection}
\acro{IEEE}{Institute of Electrical and Electronics Engineers}
\acro{ICES}{IEEE international committee on electromagnetic safety}
\acro{IEC}{international Electrotechnical Commission}
\acro{IARC}{International Agency on Research on Cancer}
\acro{IS-95}{Interim Standard 95}

\acro{KPI}{Key Performance Indicator}

\acro{LEO}{low earth orbit}
\acro{LF}{likelihood function}
\acro{LLF}{log-likelihood function}
\acro{LLR}{log-likelihood ratio}
\acro{LLRT}{log-likelihood ratio test}
\acro{LoS}{line-of-sight}
\acro{NLoS}{non-line-of-sight}
\acro{LRT}{likelihood ratio test}
\acro{wlric}[LWRIC]{lower weak restricted isometry constant}
\acro{wlrict}[LWRICt]{LWRIC threshold}
\acro{LPWAN}{low power wide area network}
\acro{LoRaWAN}{Low power long Range Wide Area Network}

\acro{MB}{multiband}
\acro{MC}{multicarrier}
\acro{MDS}{mixed distributed source}
\acro{MF}{matched filter}
\acro{m.g.f.}{moment generating function}
\acro{MI}{mutual information}
\acro{MIMO}{multiple-input multiple-output}
\acro{MISO}{multiple-input single-output}
\acrodef{maxs}[MJSO]{maximum joint support cardinality}                       
\acro{ML}[ML]{maximum likelihood}
\acro{MMSE}{minimum mean-square error}
\acro{MMV}{multiple measurement vectors}
\acrodef{MOS}{model order selection}
\acro{M-PSK}[${M}$-PSK]{$M$-ary phase shift keying}                       
\acro{M-APSK}[${M}$-PSK]{$M$-ary asymmetric PSK} 

\acro{M-QAM}[$M$-QAM]{$M$-ary quadrature amplitude modulation}
\acro{MRC}{maximal ratio combiner}                  
\acro{maxs}[MSO]{maximum sparsity order}                                      
\acro{M2M}{machine to machine}                                                
\acro{MUI}{multi-user interference}
\acro{mMTC}{massive Machine Type Communications}      
\acro{mm-Wave}[mm-Wave]{millimeter-wave}
\acro{MP}{mobile phone}
\acro{MPE}{maximum permissible exposure}
\acro{MAC}{media access control}
\acro{NB}{narrowband}
\acro{NBI}{narrowband interference}
\acro{NLA}{nonlinear sparse approximation}
\acro{NTIA}{National Telecommunications and Information Administration}
\acro{NTP}{National Toxicology Program}

\acro{OC}{optimum combining}                             
\acro{OC}{optimum combining}
\acro{ODE}{operational distortion-energy}
\acro{ODR}{operational distortion-rate}
\acro{OFDM}{orthogonal frequency-division multiplexing}
\acro{omp}[OMP]{orthogonal matching pursuit}
\acro{OSMP}[OSMP]{orthogonal subspace matching pursuit}
\acro{OQAM}{offset quadrature amplitude modulation}
\acro{OQPSK}{offset QPSK}
\acro{OFDMA}{Orthogonal Frequency-division Multiple Access}

\acro{OQPSK/PM}{OQPSK with phase modulation}

\acro{PAM}{pulse amplitude modulation}
\acro{PAR}{peak-to-average ratio}
\acrodef{pdf}[PDF]{probability density function}                      
\acro{PDF}{probability density function}
\acrodef{p.d.f.}[PDF]{probability distribution function}
\acro{PDP}{power dispersion profile}
\acro{PMF}{probability mass function}                             
\acrodef{p.m.f.}[PMF]{probability mass function}
\acro{PN}{pseudo-noise}
\acro{PPM}{pulse position modulation}
\acro{PRake}{Partial Rake}
\acro{PSD}{power spectral density}
\acro{PSEP}{pairwise synchronization error probability}
\acro{PSK}{phase shift keying}
\acro{PD}{power density}
\acro{8-PSK}[$8$-PSK]{$8$-phase shift keying}
\acro{PHP}{Poisson hole process}
 \acro{PPP}{Poisson point process}
\acrodefplural{PPP}{Poisson point processes}
\acro{FSK}{frequency shift keying}

\acro{BPP}{binomial point process}

\acro{QAM}{Quadrature Amplitude Modulation}
\acro{QPSK}{quadrature phase shift keying}
\acro{OQPSK/PM}{OQPSK with phase modulator }

\acro{RD}[RD]{raw data}
\acro{RDL}{"random data limit"}
\acro{ric}[RIC]{restricted isometry constant}
\acro{rict}[RICt]{restricted isometry constant threshold}
\acro{rip}[RIP]{restricted isometry property}
\acro{ROC}{receiver operating characteristic}
\acro{rq}[RQ]{Raleigh quotient}
\acro{RS}[RS]{Reed-Solomon}
\acro{RSC}[RSSC]{RS based source coding}
\acro{r.v.}{random variable}                               
\acro{R.V.}{random vector}
\acro{RMS}{root mean square}
\acro{RFR}{radiofrequency radiation}
\acro{RIS}{reconfigurable intelligent surface}

\acro{SA}[SA-Music]{subspace-augmented MUSIC with OSMP}
\acro{SCBSES}[SCBSES]{Source Compression Based Syndrome Encoding Scheme}
\acro{SCM}{sample covariance matrix}
\acro{SEP}{symbol error probability}
\acro{SIMO}{single-input multiple-output}
\acro{SINR}{signal-to-interference-plus-noise ratio}
\acro{SIR}{signal-to-interference ratio}
\acro{SISO}{single-input single-output}
\acro{SMV}{single measurement vector}
\acro{SNR}[\textrm{SNR}]{signal-to-noise ratio} 
\acro{sp}[SP]{subspace pursuit}
\acro{SS}{spread spectrum}
\acro{SW}{sync word}
\acro{SAR}{specific absorption rate}
\acro{SSB}{synchronization signal block}
\acro{SDG}{Sustainable Development Goal}
\acro{SBF}{spatial bandpass filtering}
\acro{SG}{stochastic geometry}

\acro{TH}{time-hopping}
\acro{ToA}{time-of-arrival}
\acro{TR}{transmitted-reference}
\acro{TW}{Tracy-Widom}
\acro{TWDT}{TW Distribution Tail}
\acro{TCM}{trellis coded modulation}
\acro{TDD}{time-division duplexing}
\acro{TDMA}{Time Division Multiple Access}

\acro{UAV}{unmanned aerial vehicle}
\acro{uric}[URIC]{upper restricted isometry constant}
\acro{urict}[URICt]{upper restricted isometry constant threshold}
\acro{UWB}{ultrawide band}
\acro{UWBcap}[UWB]{Ultrawide band}   
\acro{URLLC}{Ultra Reliable Low Latency Communications}
         
\acro{wuric}[UWRIC]{upper weak restricted isometry constant}
\acro{wurict}[UWRICt]{UWRIC threshold}                
\acro{UE}{user equipment}
\acro{UL}{uplink}

\acro{ULA}{uniform linear array}
\acro{UPA}{uniform planar array}

\acro{WiM}[WiM]{weigh-in-motion}
\acro{WLAN}{wireless local area network}

\acro{wm}[WM]{Wishart matrix}                               
\acroplural{wm}[WM]{Wishart matrices}
\acro{WMAN}{wireless metropolitan area network}
\acro{WPAN}{wireless personal area network}
\acro{wric}[WRIC]{weak restricted isometry constant}
\acro{wrict}[WRICt]{weak restricted isometry constant thresholds}
\acro{wrip}[WRIP]{weak restricted isometry property}
\acro{WSN}{wireless sensor network}                        
\acro{WSS}{wide-sense stationary}
\acro{WHO}{World Health Organization}
\acro{WP}{work package}

\acro{sss}[SpaSoSEnc]{sparse source syndrome encoding}
\acro{SO}{strategic objective}

\acro{VLC}{visible light communication}
\acro{RF}{radio frequency}
\acro{FSO}{free space optics}
\acro{IoST}{Internet of space things}

\acro{GSM}{Global System for Mobile Communications}
\acro{2G}{second-generation cellular network}
\acro{3G}{third-generation cellular network}
\acro{4G}{fourth-generation cellular network}
\acro{5G}{5th-generation cellular network}	
\acro{gNB}{next generation node B base station}
\acro{NR}{New Radio}
\acro{UN}{United Nations}
\acro{UMTS}{Universal Mobile Telecommunications Service}
\acro{LTE}{Long Term Evolution}

\acro{QoS}{Quality of Service}
\acro{ELAA}{extremely large antenna array}
\acro{CP}{coverage probability}
\acro{SE}{spectrum efficiency}
\acro{ASE}{area spectrum efficiency}
\acro{NF}{near-field}
\acro{FF}{far-field}
\acro{THz}{terahertz}
\acro{mmWave}{millimeter wave}
\end{acronym}

\begin{document}
\graphicspath{{./Figure/}}

\title{
Near-Field Multi-User Communications via Polar-Domain Beamfocusing: Analytical Framework and Performance Analysis
}
\author{Lin Chen,~\IEEEmembership{Graduate Student Member,~IEEE}, Ahmed~Elzanaty,~\IEEEmembership{Senior Member,~IEEE}, Mustafa~A.~Kishk,~\IEEEmembership{Member,~IEEE}, and Ying-Jun Angela Zhang,~\IEEEmembership{Fellow,~IEEE} 
\thanks{Lin Chen and Ying-Jun Angela Zhang are with the Department of Information Engineering, The Chinese University of Hong Kong (CUHK), Hong Kong (e-mail: \{cl022,yjzhang\}@ie.cuhk.edu.hk). }
\thanks{A. Elzanaty is with the 5GIC \& 6GIC, Institute for Communication Systems (ICS), University of Surrey, Guildford, GU2 7XH, United Kingdom (email: a.elzanaty@surrey.ac.uk).}
\thanks{M. A. Kishk is with the Department of Electronic Engineering, Maynooth University, Maynooth, W23 F2H6, Ireland (email: mustafa.kishk@mu.ie).}
}

\maketitle

\vspace{-1.5em} 

\begin{abstract}
As wireless systems evolve toward higher frequencies and extremely large antenna arrays, near-field (NF) propagation becomes increasingly dominant. Unlike far-field (FF) communication, which relies on the planar-wavefront model and is limited to angular-domain beamsteering, NF propagation exhibits spherical wavefronts that enable beamfocusing in both angle and distance, i.e., the polar domain, offering new opportunities for spatial multiple access. This paper develops an analytical stochastic geometry (SG) framework for a multi-user system assisted by polar-domain beamfocusing, which captures both NF propagation characteristics and the spatial randomness of user locations. The intrinsic coupling between angle and distance in the NF antenna pattern renders inter-user interference characterization analytically intractable. 
\color{black}
To address this challenge, we propose a tractable near-field multi-level antenna pattern (NF-MLAP) approximation, which enables computationally efficient expressions for key performance metrics.
We further establish scaling laws that characterize the dependence of area spectrum efficiency (ASE) on system parameters, such as the number of antennas and the number of simultaneously served users, revealing a fundamental trade-off between spatial multi-access gain and inter-user interference.
Numerical results validate the derived scaling laws and demonstrate that the proposed framework accurately captures performance trends, thereby providing useful design insights for future multi-user systems.
\color{black}
\end{abstract} 


\acresetall

 \vspace{-3mm}
 
\section{Introduction}\label{sec:intro}

The ever-growing demand for higher data rates has driven the adoption of \acp{ELAA} at \acp{BS} operating in high-frequency bands~\cite{6G_freq_antenna,Millimeter_tutorial}, such as \ac{mmWave}~\cite{rappaport2013millimeter}.
However, the combination of large array apertures and short wavelengths greatly extends the Rayleigh distance~\cite{selvan2017fraunhofer}, often to several hundred meters. Consequently, in a typical \ac{BS} cell,  most users are located within the {\em radiative \ac{NF} region}~\cite{10638526,BWBD,NFtutorial}. In this region, the {\em spherical-wavefront} model provides a more accurate and general representation of the electromagnetic propagation, inherently capturing \ac{NF} characteristics, while naturally reducing to the conventional {\em planar-wavefront} model of the {\em \ac{FF}} as a special case.

In conventional \ac{FF} systems, the electromagnetic wavefronts can be well approximated as planar, allowing analog beamforming to be performed primarily in the {\em angular domain} via {\em beamsteering}~\cite{NFtutorial,NFtutorial_Dai}. 
By steering narrow beams toward distinct directions, a \ac{BS} can exploit the angular-domain spatial resource to serve multiple users simultaneously. The corresponding {antenna or beam pattern}~\cite{ActualAntenna,chen2022dedicating}, characterized by a {\em mainlobe} (directed toward the intended user) and {\em sidelobes} (representing power leakage toward other directions), serves as a fundamental tool for analyzing {beamforming gain} and {inter-user interference} in FF systems~\cite{FT-MU-sector}.
 
In contrast, NF propagation exhibits spherical wavefronts, where analog beamforming transitions from angular-domain beamsteering to {\em polar-domain beamfocusing}~\cite{NFtutorial,NFtutorial_Dai}. In this case, both angle and distance are jointly exploited to concentrate beam energy at a specific spatial point~\cite{chen2025quasi}.
This additional spatial dimension enables polar-domain multiple access, where users can be distinguished and simultaneously served based on both their angles and distances~\cite{kosasih2024finite,LDMA}.  
Despite the promise of enhancing spatial resource reuse, beamfocusing does not eliminate {\em inter-user interference}. A beam intended for one user may still leak power to nearby users, especially when user locations are densely distributed.
Unlike the FF antenna pattern, which varies only with angle, the {\em NF antenna pattern} is intrinsically coupled in both {\em angle and distance}~\cite{kosasih2024finite,BWBD}, making interference characteristics highly dependent on the {\em spatial randomness of user locations}. As a result, analyzing inter-user interference in the NF is considerably more challenging than that in the FF.

These fundamental distinctions between NF and FF propagation necessitate a new analytical framework that can accurately characterize NF channels, polar-domain beamfocusing, and interference under the spatial randomness of user locations. Such a framework is essential for evaluating the performance of NF multi-user networks and for understanding the key performance trade-offs, e.g., between spatial multi-access gain and inter-user interference, thereby offering valuable design insights for future multi-user systems. 

\vspace{-2mm}

\subsection{Related Works}\label{subsec:related}

Recent research on \ac{NF} communications with \acp{ELAA} has primarily focused on channel modeling, channel estimation, and beamforming strategies~\cite{NFtutorial,NFtutorial_Dai}.  
Ray-tracing is a widely used technique for modeling wireless channels. By explicitly accounting for the positions of the transmitter, receiver, and scatterers, ray-tracing captures the multipath propagation of electromagnetic waves, including both amplitude variations and phase shifts across the antenna array. 
In the NF, the ray-tracing model is often approximated using the uniform spherical-wavefront model~\cite{USW,zhang2022fast,cui2022channel}, where the array response vector exhibits constant amplitude, while the phase shifts vary with both the angle and distance of a source or target. This Fresnel approximation allows the NF channel to be efficiently parameterized using low-dimensional angular and distance information~\cite{cui2022channel,xu2024near},  directly linking the channel state information to the users' locations.
Therefore, parametric channel estimation and localization-based methods have emerged as efficient NF channel acquisition schemes~\cite{kosasih2023parametric,yuan2024scalable,11150585,11016742,chen2025tracking}. 
 
The NF array response vector physically characterizes the amplitude-phase profile across the array induced by a wave originating from a specific point in the NF.
Hence, it can be used to construct an analog beamforming vector that concentrates energy at a {\em desired (focal) point}, i.e., a beamfocusing vector~\cite{NFtutorial,chen2025quasi}.
The \ac{NF} antenna pattern, defined as the squared inner product of beamfocusing vectors, describes the focusing gain at the focal point and the power leakage elsewhere.
As analyzed in \cite{BWBD,kosasih2024finite},  beam energy is concentrated around the focal point with finite ranges in both distance and angle, termed the {\em beam depth and width}. 
This property makes polar-domain beamfocusing a promising enabler for multiple access.
Specifically, in a multi-user system, after obtaining the user locations, the BS can serve multiple users simultaneously by generating focused beams tailored to distinct locations~\cite{kosasih2024finite,LDMA}.  
Importantly, these beamfocusing vectors with constant-amplitude phase-varying elements can be implemented using low-cost analog phase shifters, making polar-domain multiple access highly practical for multi-user systems even with a limited number of \ac{RF} chains.  

Based on Monte-Carlo simulations, the authors in \cite{kosasih2024finite} evaluated the sum rate of a NF multi-user network, where an ELAA-assisted BS serves 2-8 single-antenna users randomly distributed in distance but aligned in angle, and demonstrated the potential of beamfocusing for distance-domain resource reuse. 
The location division multiple access scheme proposed in \cite{LDMA} exploited both angle and distance separability in NF (and only angle separability in FF) for multi-user communications. It was shown that the inter-user interference is generally non-negligible under random user separation (e.g., 10 users uniformly distributed in a 120-degree sector), but can be effectively reduced through careful user placement according to the beam depth and width.

Despite providing useful insights, the above works typically consider deterministic user placements and rely on Monte-Carlo simulations.
Such approaches lack analytical tractability, limiting their ability to reveal system-level scaling laws and fundamental design trade-offs.
This gap motivates the development of \ac{SG}-based analytical frameworks for NF multi-user networks, enabling scalable performance characterization and deeper theoretical understanding.

In general, \ac{SG}~\cite{SGtutorial,andrews2016primer} provides a powerful analytical tool for modeling and analyzing wireless systems under spatial randomness, enabling tractable derivation of performance metrics such as \ac{CP}, \ac{SE}, and \ac{ASE}. 
However, conventional SG frameworks designed for FF systems, e.g.,~\cite{multi-cos,FlatTop,chen2024coverage,FT-SU-MC, chen2024joint,DiscreteAntenna}, are inadequate for NF analysis. 
First, conventional SG analyses typically assume users' locations follow a \ac{PPP} over an infinite area, resulting in a random and potentially unbounded number of users. In a practical multi-user system, the BS simultaneously serves a finite number of users within a cell/sector, constrained by the available \ac{RF} chains and service region.
This user randomness within a finite region is more accurately captured by a \ac{BPP}~\cite{BPP1,BPP2}.   
More fundamentally, the NF antenna pattern depends jointly on both angle and distance, making the analytical characterization of inter-user interference far more complex than that in the FF, where only angular dependence exists. The lack of tractable closed-form expressions for the NF antenna pattern further complicates this analysis.

Overall, a practical and tractable SG framework for NF polar-domain beamfocusing assisted multi-user systems has so far received very limited attention. 
An analytical framework that accurately reflects NF propagation characteristics, captures user randomness within a finite area, and provides a tractable characterization of inter-user interference is still lacking. This gap leaves the trade-offs between hardware configuration (including the number of antennas and \ac{RF} chains) and system performance (in terms of spatial resource reuse and interference mitigation) poorly understood. 
 
\vspace{-2mm}

\subsection{Contributions}\label{subsec:contri}

This paper addresses the lack of analytical tools for evaluating polar-domain multiple access in ELAA-assisted high-frequency systems dominated by NF propagation. 
Leveraging tools from \ac{SG}, we model and analyze a multi-user, single-cell network with sectorization, where a \ac{BS} equipped with an ELAA and a limited number of RF chains simultaneously serves multiple users in a sector via analog beamfocusing. 
The main contributions are summarized as follows:
\begin{itemize}
    \item We develop a general \ac{SG} framework for NF multi-user communications assisted by polar-domain beamfocusing in a high-frequency system, explicitly capturing NF propagation characteristics and user spatial randomness. Within this framework, we derive exact analytical expressions for key performance metrics, including  \ac{CP}, \ac{SE}, and \ac{ASE}, based on the exact NF antenna pattern. 

    \item To achieve analytical tractability, we propose the near-field multi-level antenna pattern (NF-MLAP).
    It discretizes the exact NF pattern into finite angular and distance levels, preserving the key spatial features of NF beamfocusing while naturally reducing to a pattern that depends only on angle in the FF region.
    \color{black}
    This approximate pattern enables tractable inter-user interference analysis.  
    Based on the NF-MLAP, we derive tractable approximations for CP, SE, and ASE. We also provide guidelines for selecting the discretization level to ensure accurate performance approximation.
    \item Building on the proposed analytical framework,  
    we establish explicit scaling laws that characterize the dependence of ASE on the number of antennas and the number of simultaneously served users. 
    The analysis reveals a fundamental trade-off between spatial reuse gain and inter-user interference, whose scaling behaviors differ across low- and high-threshold regimes.
    These findings are validated by numerical results, demonstrating that the proposed framework accurately captures performance trends and provides useful guidance for system design.
    \color{black}
\end{itemize}

		{\em Organization:} The remainder of this paper is organized as follows. Sec.~\ref{sec:sys} provides the SG-based spatial model, presents the NF-based signal model, and introduces the performance metrics. Sec.~\ref{sec:analy} derives the analytical expressions for the performance metrics in a general form. Sec.~\ref{sec:approx_pattern} proposes the approximate NF antenna pattern. Sec.~\ref{sec:approx_analy} provides tractable approximations of performance metrics.  \textcolor{black}{Sec.~\ref{sec:scaling_insight} reveals how system performance scales with the key parameters.}
        Sec.~\ref{sec:results} shows the numerical results. Sec.~\ref{sec:conclusion} concludes this work \textcolor{black}{and discusses future research directions.}

		{\em Notations:} 
		The upper- and lower-case bold letters denote matrices and vectors, respectively. 
		For a matrix or vector, $\mathbf{(\cdot)}^{\mathsf{T}}$, $\mathbf{(\cdot)}^{*}$, and $\mathbf{(\cdot)}^{\mathsf{H}}$ represent transpose, conjugate, and conjugate transpose operators, respectively.  $\left\|\cdot\right\|_2$ denotes the Euclidean norm. $\mathbf{I}$ denotes the identity matrix of an appropriate size. $\mathcal{CN}(\mathbf{0},\mathbf{I})$ denotes a standard complex Gaussian distribution.
        Moreover,  ${\rm j}$ is the imaginary unit with ${\rm j}^2=-1$.  
        For real numbers $a \in \mathbb{R}$ and $b\in \mathbb{R}$, ${C}^{a}_{b}$ represents the binomial coefficient, and 
		$\left \lfloor a \right \rfloor $ represents the floor function.  
        For a function $F(x)$, $F(x)|_{a}^{b}=F(b)-F(a)$. 
        The symbol $\mathcal{O}(\cdot)$ denotes the standard big-O notation, which characterizes the asymptotic order.
        A summary of other notations used in this paper is provided in Table~\ref{tab:TableOfNotations}. 


\begin{table}[t]\caption{Table of notations.}
\vspace{-3mm}
\centering
\begin{center}
{ \linespread{1}
\renewcommand{\arraystretch}{1.3}
    \begin{tabular}{ {c} | {c} }
    \hline
        \hline
    \textbf{Notation} & \textbf{Description} \\ \hline
    $N$  & Antenna number ($N=2\bar N+1$).\\ \hline
    $n$  & Index of an antenna, $n\in \{-\bar N,...,\bar N\}$. \\ \hline
    $N_{\rm s}$  & Number of sectors in a cell.\\ \hline
    $\Psi_{\rm a}$  & BPP modeling the locations of active users. \\ \hline
    $N_{\rm a}$  & Number of active users (or RF chains) within a sector.\\ \hline
    $\kappa$ or $\kappa'$ & Index of an active user, $\kappa,\kappa' \in \{1,...,N_{\rm a}\}$. \\ \hline
    $\mathbf{u}_\kappa\!=\![\theta_\kappa,r_\kappa]^{\mathsf{T}}$ & Polar coordinate of the $\kappa$-th active user location. \\ \hline
    $\theta;\vartheta$ &  Physical angle; spatial angle ($\vartheta=\frac{1}{2}\sin\theta$). \\ \hline
    $R_{\rm c}$; $R_{\rm L}$ & Radius of a cell; a LoS ball.\\ \hline
    $r_{\kappa,n}$ & Distance from the $\kappa$-th active user to the $n$-th antenna.\\ \hline
    $\lambda$ & Wavelength ($\lambda$) of the center carrier frequency ($f_{\rm c}$). \\ \hline
    $\zeta$ & Reference path loss $\zeta=({\lambda}/{4 \pi })^2$.  \\ \hline
    $\alpha$ & Path-loss exponent.  \\ \hline 
    $z$; $\sigma^2$ & Additive noise $z$ with power $\sigma^2$\\ \hline
    $P_{\rm t}$ & Total transmit power of a BS.  \\ \hline 
    $\mathcal{G} (\cdot)$ & {Antenna pattern} \\ \hline
    $\mathbf{h}; \mathbf{a} ; \mathbf{w}$ & Channel; array response vector; beamforming vector. \\ \hline 
    $s_{\kappa}$ & Transmitted symbol for the $\kappa$-th active user.\\ \hline 
    $\mathbf{s}$ & $\mathbf{s}=[s_{1}, ..., s_{N_{\rm a}}]^\mathsf{T}$ with $\mathbf{s}\sim\mathcal{CN}(\mathbf{0},\mathbf{I})$.\\ \hline
    $y_\kappa$ & Received signal at the $\kappa$-th active user. \\ \hline 
    $y_{\kappa,\rm S}$ or $y_{\kappa,\rm I}$  & Received serving or interfering signal at $\mathbf{u}_\kappa$. \\ \hline 
    $\tau$ & Predefined threshold. \\ \hline  
    $F_X(x)$, $f_X(x)$ &
    CDF and PDF of $X$.\\ \hline
    \end{tabular}} 
\end{center}
\label{tab:TableOfNotations}
\end{table}

\vspace{-2mm}

\section{System Model}\label{sec:sys}

In this section, we first present the \ac{SG}-based spatial model of a multi-user single-cell network operating in a high-frequency band. We then provide the signal model and introduce performance metrics.

\begin{figure}[t!]
\centering
\includegraphics[width=0.66\columnwidth]{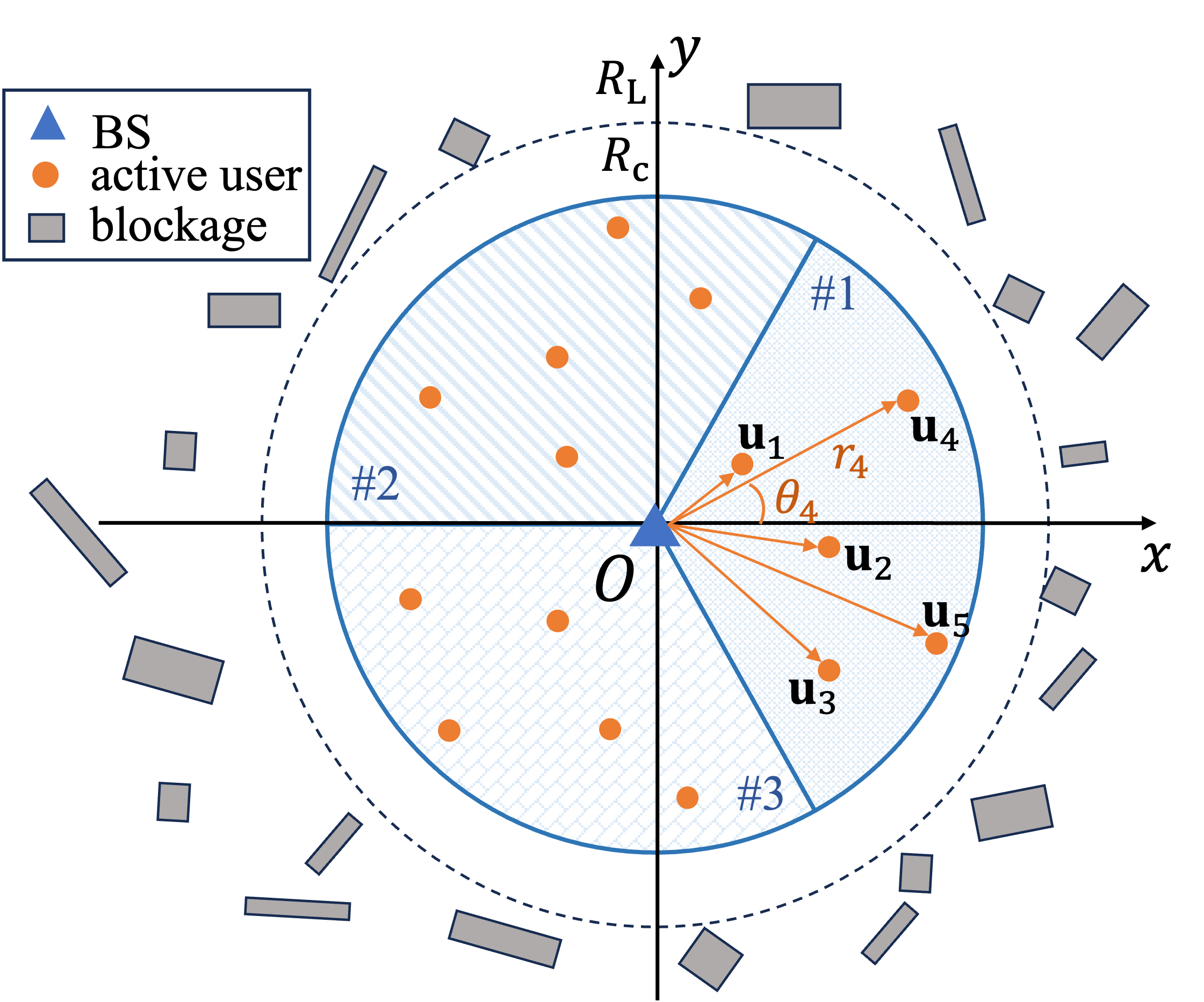}
\vspace{-1mm}
\caption{Spatial modeling of a multi-user single-cell system with three sectors, where a \ac{BPP} models the random locations of active users and a \ac{LoS} ball models the blockage effect.} 
\label{fig:sys}
\end{figure}

\vspace{-2mm}

\subsection{Spatial Model}\label{subsec:network}

We consider high-frequency downlink transmission in a single cell, consisting of multiple single-antenna users and a multi-antenna \ac{BS}, as shown in Fig.~\ref{fig:sys}. 
The \ac{BS} is located at the origin, and the cell radius is denoted as $R_{\rm c}$. 
\color{black}
To capture blockage effects, we adopt a \ac{LoS} ball model~\cite{FlatTop,ActualAntenna,Millimeter_tutorial} with radius $R_{\rm L}$, where a link between a user and the BS is \ac{LoS} if their distance satisfies $r \leq R_{\rm L}$.
This approximation is widely used in \ac{SG}-based system performance analysis, as it provides a good balance between analytical tractability and modeling accuracy~\cite{FlatTop,ActualAntenna,Millimeter_tutorial}. 
Since high-frequency signals (e.g., \ac{mmWave}) suffer from severe path loss and are highly susceptible to blockages, their propagation is generally dominated by short-range \ac{LoS} links~\cite {Millimeter_tutorial,indoorSM}. 
Accordingly, we focus on a \ac{LoS}-dominant scenario and assume the cell lies within the \ac{LoS} ball, i.e., $R_{\rm c} \le R_{\rm L}$.
\color{black}

To support spatial multiple access, the cell is uniformly divided into $N_{\rm s}$ sectors, each covering an angular width of $\frac{2\pi}{N_{\rm s}} $. A typical choice is $N_{\rm s}=3$~\cite{rebato2018multi,tse2005fundamentals}. 
Moreover, the \ac{BS} is equipped with $N_{\rm s}$
\acp{ULA}, each dedicated to one sector and consisting of $N$ antennas.
Without loss of generality, our analysis focuses on a representative sector spanning the angular range $[-\frac{\pi}{N_{\rm s}},\frac{\pi}{N_{\rm s}}]$.  The \ac{BS} allocates $N_{\rm a}$ limited \ac{RF} chains to this sector, where $N_{\rm a}\ll N$.

\color{black}
Consider a fully loaded case where the number of potential users in the sector exceeds the number of available \ac{RF} chains. Therefore, in each time-frequency resource block, the BS selects and serves $N_{\rm a}$ users simultaneously, referred to as {\em active users}.
Note that the spatial distribution of active users depends on both the underlying potential-user distribution and the user selection mechanism. 
We model the active users as a \ac{BPP}~\cite{BPP1,BPP2}, which is suitable for modeling a fixed number of nodes in a finite region and thus captures a representative active-user distribution for the subsequent analysis. We denote the \ac{BPP} as $\Psi_{\rm a}$, where $N_{\rm a}$ active users are i.i.d. distributed within the sector. 
\color{black}
For clarity, we order the active users by their distance from the BS. Specifically, the $\kappa$-th active user is located at $\mathbf{u}_{\kappa}=[\theta_\kappa,r_\kappa]^{\mathsf{T}} \in \Psi_{\rm a}$, where $r_\kappa\le r_{\kappa'}$ if $\kappa\le \kappa'$ for $\kappa,\kappa'\in \{1,...,N_{\rm a}\}$. Unless stated otherwise, we use polar coordinates in the following.
\textcolor{black}{Moreover, we refer to the $\kappa$-th active user as the {\em tagged user} in the subsequent analysis.}

\vspace{-2mm}

\subsection{Signal Model}\label{sec:signal}
\subsubsection{Received Signal}
We denote the downlink channel between the $\kappa$-th active user and the BS by $\mathbf{h}^{\mathsf{H}}_{\kappa} \in \mathbb{C}^{1\times N}$, 
the beamforming matrix by $\mathbf{W}=[\mathbf{w}_1, ..., \mathbf{w}_{N_{\rm a}}] \in \mathbb{C}^{N\times N_{\rm a}}$ with $\left\|\mathbf{w}_{\kappa}\right\|_2^2=1$, and the symbol for the $\kappa$-th active user by $s_{\kappa}$. 
We assume that the transmitted symbols are independent and identically distributed (i.i.d.), each with zero mean and unit power, i.e.,  $\mathbf{s}=[s_{1}, ..., s_{N_{\rm a}}]^\mathsf{T}$ and $\mathbf{s}\sim\mathcal{CN}(\mathbf{0},\mathbf{I})$.
The total transmit power $P_{\rm t}$ is equally allocated across $N_{\rm a}$ active users. 
\color{black}
Since multiple beams are transmitted simultaneously, the beam intended for one active user contributes to interference at another active user.
From the perspective of the tagged user located at $\mathbf{u}_\kappa$, the beams intended for other active users located at $\mathbf{u}_{\kappa'}\in \Psi_{\rm a}\setminus\{\mathbf{u}_{\kappa}\}$ cause interference at the tagged user. Accordingly, we treat other active users as {\em interferers}, each associated with one transmit beam. 
\color{black} 
Therefore, the received signal at the $\kappa$-th active user, denoted by $y_{\kappa}$, consists of a serving signal from the intended beam, denoted by $y_{\kappa,\rm S}$, and an interfering signal from other beams, denoted by $y_{\kappa,\rm I}$.
Mathematically, 
   \begin{align} \label{eq:y}
 	y_{\kappa}  & 
    = \underbrace{\sqrt{\frac{P_{\rm t}}{ N_{\rm a} } }  \mathbf{h}_{\kappa}^{\mathsf{H}} \mathbf{w}_{\kappa} s_{\kappa}}
_{y_{\kappa,\rm S}}  +  \underbrace{\sqrt{\frac{P_{\rm t}}{ N_{\rm a} } } \sum_{\kappa'\neq\kappa} \mathbf{h}_{\kappa}^{\mathsf{H}} \mathbf{w}_{\kappa'}  s_{\kappa'}}
_{y_{\kappa,\rm I}}
+ {z}_{\kappa},
 \end{align} 
where ${z}_{\kappa}\sim \mathcal{CN}(0,\sigma^2)$ is the additive noise with power $\sigma^2$.

\subsubsection{Channel Model}\label{subsec:channel}
\color{black}
\textcolor{black}{In the considered LoS scenario, the high-frequency channel is typically dominated by the LoS component, while the \ac{NLoS} component is generally much weaker than the \ac{LoS} component~\cite{ActualAntenna,FlatTop,20dB}. 
Moreover, the small-scale fading on the LoS path has a negligible impact on the system performance~\cite{FlatTop,rappaport2013millimeter}.
Accordingly, we adopt a simplified channel model~\cite{zhang2022fast} that explicitly captures the dominant \ac{LoS} component, while neglecting the weak \ac{NLoS} component and the small-scale fading on the LoS path.
The validity of this modeling assumption will be further justified numerically in Sec.~\ref{subsec:fading-NLoS}.}
\color{black}

For notational simplicity, we assume that the antenna number of the \ac{ULA} is odd with $N=2\bar{N}+1$, indexed as $n \in \{-\bar N, ...,0,..., \bar N\}$. The antenna spacing is $d=\frac{\lambda}{2}$, where $\lambda$ is the wavelength associated with the carrier frequency $f_{\rm c}$. The array aperture is $D=(N-1)d$.
Under the \ac{NF} spherical-wavefront model, the LoS channel between the BS and the $\kappa$-th active user located at $\mathbf{u}_\kappa=[\theta_{\kappa},r_\kappa]^{\mathsf{T}}$ is given by~\cite{zhang2022fast} 
\begin{align}\label{eq:channel}
\mathbf{h}_{\kappa}^{\mathsf{H}}=\sqrt{\zeta r_\kappa^{-\alpha} }{\rm e}^{-{\rm j}\frac{2\pi}{\lambda} r_\kappa}\sqrt{N}\mathbf{a}^{\mathsf{H}}(\theta_{\kappa},r_\kappa).
\end{align}
Here, $\zeta=(\frac{\lambda}{4 \pi })^2$ is the reference path loss, $\alpha$ is the path loss exponent, 
and $\mathbf{a}^{\mathsf{H}}(\theta_{\kappa},r_\kappa) $ denotes the \ac{NF} array response vector, which is given by
\begin{align}\label{eq:array}
    & \!\mathbf{a}^{\mathsf{H}}(\theta_{\kappa},r_\kappa) \!= \!\frac{1}{\sqrt{N}}\left[{\rm e}^{-{\rm j}\frac{2\pi}{\lambda}(r_{\kappa,-\bar {\small N}}-r_\kappa)}, ..., {\rm e}^{-{\rm j}\frac{2\pi}{\lambda} (r_{\kappa,\bar {\small N}} -r_\kappa)}\right]\!,\!
\end{align} 
where $r_{\kappa,n}=\sqrt{r_\kappa^2+n^2d^2-2r_\kappa n d \sin \theta_{\kappa} }$
is the distance between the $\kappa$-th active user and the $n$-th antenna.

Unlike the classic FF planar-wavefront model, where the phase shift $\frac{2\pi}{\lambda} (r_{\kappa,n} -r_\kappa)$ in the $n$-th antenna can be accurately approximated using the {\em first-order} Taylor expansion of $r_{\kappa,n}$ (i.e., {\em Fraunhofer approximation}), the NF spherical-wavefront model requires retaining the {\em second-order} term to capture the wavefront curvature (i.e., {\em Fresnel approximation})~\cite{selvan2017fraunhofer,NFtutorial}. Specifically, applying the second-order Taylor expansion $\sqrt{1+x}\approx 1+\frac{1}{2}x-\frac{1}{8}x^2$ to $r_{\kappa,n}$ yields~\cite{cui2022channel}
\begin{align}
\begin{split}\label{eq:rNF}
    r_{\kappa,n}\!-r_\kappa&\approx 
    -nd \sin\theta_{\kappa} +n^2d^2\frac{1-\sin^2\theta_{\kappa}}{2r_\kappa},   
\end{split}
\end{align}
where the quadratic term reflects the NF spherical wavefront, making the phase shifts across antennas dependent on both angle $\theta_\kappa$ and distance $r_\kappa$.

Notably, the \ac{NF} model incorporates the FF model as a special case.
When the array aperture is small, or the user is sufficiently far from the BS, the quadratic term in \eqref{eq:rNF} becomes negligible, reducing $r_{\kappa,n}-r_\kappa$ to $-nd\sin\theta_\kappa$. This corresponds to the classical Fraunhofer approximation. In this FF regime, $\mathbf{a}^{\mathsf{H}}(\theta_{\kappa},r_\kappa)$ simplifies to the {\em \ac{FF} array response vector} as
\begin{align}\label{eq:aA}
    \mathbf{a}_{\rm A}^{\mathsf{H}}(\theta_\kappa)\!=\! 
\frac{1}{\sqrt{N}} [{\rm e}^{-{\rm j}\frac{2\pi}{\lambda}{\bar N}d \sin\theta_{\kappa} }, ...,   
  {\rm e}^{{\rm j}\frac{2\pi}{\lambda} {\bar N}d \sin\theta_{\kappa}   }],
\end{align}
where the phase shifts depend only on the angle $\theta_\kappa$.

\subsubsection{Analog Beamfocusing and Received Power}\label{subsec:analog}
\color{black}
Consider that the spatial locations of active users are available at the BS, which can be obtained via efficient estimation, localization, or tracking techniques~\cite{kosasih2023parametric,yuan2024scalable,11150585,11016742,xu2024near,chen2025tracking}.
Leveraging this location information, the BS employs analog beamforming through matched filtering, i.e., the beamforming vector for the $\kappa$-th active user is set as
$\mathbf{w}_{\kappa}=\mathbf{a}(  \theta_{\kappa},  r_\kappa)$.
This is a tractable and practical approach in large-antenna systems with a limited number of \ac{RF} chains, as $\mathbf{a}(  \theta_{\kappa},  r_\kappa)$  with constant-magnitude elements can be implemented using low-cost analog phase shifters.
\color{black}
Correspondingly, the power of the received serving signal $y_{\kappa,\rm S}$ in \eqref{eq:y}, 
denoted by $P_{\kappa,{\rm S}}=\mathbb{E}_{\mathbf{s}}[|y_{\kappa,\rm S}|^2]$, is 
\begin{align}\label{eq:yServing}
P_{\kappa,{\rm S}} &= \frac{P_{\rm t}}{N_{\rm a}}  |\mathbf{h}_{\kappa}^{\mathsf{H}}\mathbf{w}_{\kappa}|^2 = \frac{P_{\rm t}}{ N_{\rm a} } \zeta r_\kappa^{-\alpha}  N|\mathbf{a}^{\mathsf{H}} (\theta_{\kappa},r_\kappa) \mathbf{a} ( \theta_{\kappa},r_\kappa)|^2 
\nonumber\\&= \frac{P_{\rm t}}{ N_{\rm a} } N\zeta r_\kappa^{-\alpha}.     
\end{align}
This indicates perfect alignment of the beam with the user's channel, concentrating energy at the user's location and achieving a beamforming gain of $N$. 
Accordingly, $\mathbf{a}(\theta_\kappa,r_\kappa)$ is termed the {\em \ac{NF} beamfocusing vector} with the focal point being $[\theta_\kappa,r_\kappa]^{\mathsf{T}}$.\footnote{When the user lies in the FF, NF beamfocusing $\mathbf{a} (\theta_{\kappa},r_\kappa)$ naturally reduces to FF beamsteering $\mathbf{a}_{\rm A} (\theta_{\kappa})$, directing beam energy along the angle $\theta_\kappa$.} 
Under the i.i.d. symbol assumption, i.e., $\mathbf{s}\sim\mathcal{CN}(\mathbf{0},\mathbf{I})$, the power of the received interfering signal $y_{\kappa,\rm I}$ in \eqref{eq:y},
denoted by $P_{\kappa,{\rm I}}=\mathbb{E}_{ \mathbf{s}}[|y_{\kappa,\rm I}|^2]$, is
\begin{align}\label{eq:yInterfering}
    P_{\kappa,{\rm I}} &= \frac{P_{\rm t}}{N_{\rm a}}  \sum_{\kappa'\neq\kappa} |\mathbf{h}_{\kappa}^{\mathsf{H}}\mathbf{w}_{\kappa'}|^2 
    \nonumber\\&= \frac{P_{\rm t}}{ N_{\rm a} }   \sum_{\kappa'\neq\kappa} \zeta r_\kappa^{-\alpha} N |\mathbf{a}^{\mathsf{H}} (\theta_{\kappa},r_\kappa) \mathbf{a} ( \theta_{\kappa'},r_{\kappa'})|^2, 
\end{align} 
\textcolor{black}{which captures the aggregate interference at the tagged user caused by the beams intended for other active users.} 
 
\begin{figure}
\vspace{-2mm}
    \centering
\includegraphics[width=0.75\linewidth]{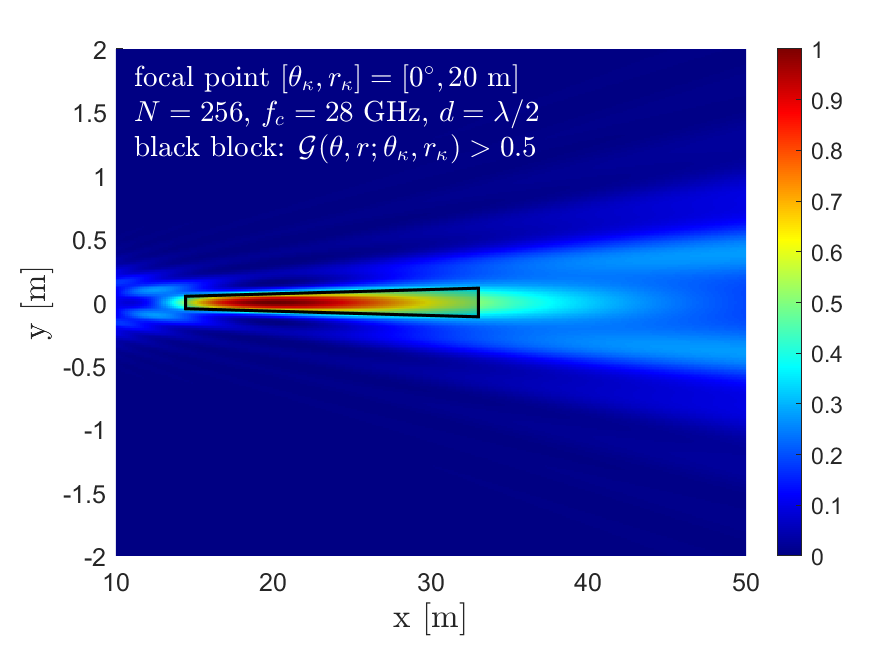}
\vspace{-3mm}
    \caption{Visualization of the NF antenna pattern, where a \ac{ULA} is aligned with the $\rm y$-axis and its center is at the origin.}
    \label{fig:pattern}
\end{figure}

\subsubsection{Antenna Pattern}\label{subsec:ActPattern}
\color{black}
Both the received serving and interfering signal powers in \eqref{eq:yServing} and \eqref{eq:yInterfering} are determined by the normalized beamforming gain, defined as 
\begin{align} 
\mathcal{G}(\theta_{\rm o},r_{\rm o};\theta_{\rm f},r_{\rm f})
\triangleq |\mathbf{a}^{\mathsf{H}}(\theta_{\rm o},r_{\rm o})\mathbf{a}(\theta_{\rm f},r_{\rm f})|^2.
\end{align}
This function characterizes how a beam focused on $[\theta_{\rm f},r_{\rm f}]^{\mathsf{T}}$ (i.e., focal point) distributes energy across any location $[\theta_{\rm o},r_{\rm o }]^{\mathsf{T}}$ (i.e., observation point).
We refer to $\mathcal{G}(\theta_{\rm o},r_{\rm o};\theta_{\rm f},r_{\rm f})$ as the {\em \ac{NF} polar-domain  antenna pattern}. 
In particular,  
$\mathcal{G}(\theta_\kappa,r_\kappa;\theta,r)$ characterizes the interference received by the tagged user at $\mathbf{u}_\kappa$ from the beam focused on another user at $\mathbf{u}=[\theta,r]^{\mathsf T}\in\Psi_{\rm a}\setminus\{\mathbf{u}_\kappa\}$. 
Note from the symmetry of the inner product that $\mathcal{G}(\theta_\kappa,r_\kappa;\theta,r)
=
\mathcal{G}(\theta,r;\theta_\kappa,r_\kappa)$.
This symmetry allows the interference experienced by the tagged user to be characterized by the spatial distribution of other active users over the polar domain using the antenna pattern $\mathcal{G}(\theta,r;\theta_\kappa,r_\kappa)$ with $[\theta_\kappa,r_\kappa]^\mathsf{T}$ as the reference point.

\color{black}
The explicit form of $\mathcal{G} (\cdot)$ under the NF model is~\cite{cui2022channel}  
\begin{align}\label{eq:antenna_pattern}
    & \mathcal{G} (\theta,r ; \theta_\kappa, r_\kappa) \overset{(a)}{=} \frac{1}{N^2} \bigg| \sum_{n=-\bar N}^{\bar N} \exp\bigg( {\rm j}\Phi_n^{\rm ang}+{\rm j}\Phi_n^{\rm dist} \bigg) \bigg|^2,
\end{align}
where $\Phi_n^{\rm ang}=\frac{2\pi}{\lambda}n d(\sin\theta-\sin\theta_\kappa)$ is a linear angle-dependent phase term, $\Phi_n^{\rm dist}=\frac{2\pi}{\lambda}(nd)^2(\frac{1-\sin^2\theta_\kappa}{2r_\kappa}-\frac{1-\sin^2\theta}{2r})$ is
a quadratic distance-dependent phase term, and (a) is from the Fresnel approximation in \eqref{eq:rNF}.  
Particularly, $\mathcal{G} (\cdot)$ peaks at the focal point with $\mathcal{G} (\theta_\kappa, r_\kappa ; \theta_\kappa, r_\kappa)=1$; when observed at other points, 
$\mathcal{G} (\theta, r ; \theta_\kappa, r_\kappa)<1$, reflecting energy leakage.
As shown in Fig.~\ref{fig:pattern}, most energy concentrates around the focal point, forming a region with limited angular and distance ranges, within which $\mathcal{G} (\theta, r; \theta_\kappa, r_\kappa)>0.5$.

Note that the NF antenna pattern incorporates the FF antenna pattern as a special case. As discussed in Sec.~\ref{subsec:channel}, as the antenna number $N$ decreases or the focal distance $r_\kappa$ increases, the \ac{NF} model gradually transitions to the \ac{FF} model.
In this limit, the NF antenna pattern $\mathcal{G}(\theta,r;\theta_\kappa,r_\kappa)$ in \eqref{eq:antenna_pattern} simplifies into the classical FF antenna pattern $\mathcal{G}_{\rm A}(\theta;\theta_\kappa)\triangleq |\mathbf{a}_{\rm A}(\theta)^{\mathsf{H}} \mathbf{a}_{\rm A}(\theta_\kappa )|^2$, i.e.,~\cite{cui2022channel,ActualAntenna} 
\begin{align}\label{eq:FF}
		\mathcal{G}_{\rm A}(\theta;\theta_\kappa) & =\frac{1}{N^2}\bigg|\sum_{n=-\bar N}^{\bar N}\exp\big({\rm j} \Phi_{n}^{\rm ang}\big)\bigg|^2
		\nonumber\\&=\bigg|\frac{\sin(\frac{\pi}{2}N (\sin\theta_\kappa-\sin\theta))}{N\sin(\frac{\pi}{2} (\sin\theta_\kappa-\sin\theta))} \bigg|^2,
\end{align}
which is solely related to the angle.
Comparing \eqref{eq:antenna_pattern} with \eqref{eq:FF}, the NF antenna pattern varies jointly with angle and distance, introducing much higher complexity than its FF counterpart.

 \vspace{-2mm}
\subsection{Performance Metrics}\label{subsec:metrics}

From the above discussion, the BS can spatially reuse polar-domain resources to simultaneously serve multiple users via NF beamfocusing, while the performance of such spatial multiple access is fundamentally constrained by inter-user interference, which is captured by the antenna pattern $\mathcal{G}(\cdot)$ in \eqref{eq:antenna_pattern}.
To quantify the impact of such interference, we start with the \ac{SIR}-related performance metrics.\footnote{The extension to \ac{SINR}-related metrics will be given in Sec.~\ref{subsec:SINR}.} 
With \eqref{eq:yServing} and \eqref{eq:yInterfering}, the \ac{SIR} of the $\kappa$-th active user is 
\begin{align}\label{eq:SIR_k}
	{\rm SIR}_{\kappa}= \frac{P_{\kappa,{\rm S}}}{P_{\kappa,{\rm I}}}   
    = \frac{|\mathbf{a}^{\mathsf{H}} (\theta_{\kappa},r_\kappa)\mathbf{a} (\theta_{\kappa}, r_\kappa)|^2 }{\sum_{\kappa'\neq\kappa} |\mathbf{a}^{\mathsf{H}} (\theta_{\kappa},r_\kappa)\mathbf{a} ( \theta_{\kappa'}, r_{\kappa'})|^2 } 
  =  \frac{1}{I_{\kappa}},\!
\end{align}
where $I_{\kappa}\triangleq \sum_{\kappa'\neq\kappa} \mathcal{G} (\theta_{\kappa},r_\kappa; \theta_{\kappa'}, r_{\kappa'})$ \textcolor{black}{represents the normalized aggregate interference from other active users.}  

A fundamental reliability metric is the \ac{CP}, which quantifies the probability that the \ac{SIR} exceeds a predefined threshold ($\tau$) required for successful communication. The {\em conditional \ac{CP}} given the $\kappa$-th active user at $[\theta_\kappa, r_\kappa]^{\mathsf{T}}$ can be expressed as 
\begin{align}\label{eq:condCP}
		{\rm CP}_{\kappa}(\tau|\theta_\kappa, r_\kappa)=\mathbb{P}\{{\rm SIR}_{\kappa}>\tau | \theta_\kappa, r_\kappa\}.
	\end{align}
Another important metric is the \ac{SE}, which quantifies the rate of successfully transmitted data for each active user. 
Given the $\kappa$-th active user at $[\theta_\kappa, r_\kappa]^{\mathsf{T}}$, the {\em conditional \ac{SE}} can be practically computed as~\cite{BPP1}
\begin{align}\label{eq:condAR}
		{\rm SE}_{\kappa} (\tau|\theta_\kappa, r_\kappa)&= \mathbb{P}\{{\rm SIR}_{\kappa}>\tau | \theta_\kappa, r_\kappa\}\log_2(1+\tau)\nonumber\\&=  {\rm CP}_{\kappa}(\tau|\theta_\kappa, r_\kappa)\log_2(1+\tau).
	\end{align}
Averaging over the spatial distribution of the $\kappa$-th active user gives the overall \ac{CP} and the overall \ac{SE}, i.e.,
\begin{align}\label{eq:overCP}
		&\!{\rm CP}_{\kappa}(\tau)\!=\!\mathbb{E}_{\theta_\kappa, r_\kappa}[{\rm CP}_{\kappa}(\tau|\theta_\kappa, r_\kappa)],
		\\&{\rm SE}_{\kappa} (\tau)\!=\!  \mathbb{E}_{\theta_\kappa, r_\kappa}[{\rm SE}_{\kappa} (\tau|\theta_\kappa, r_\kappa)] \!=\! {\rm CP}_{\kappa}(\tau)\log_2(1+\tau) . \label{eq:overAR}
	\end{align}
To evaluate the overall sector efficiency of multi-user transmission, we further adopt the \ac{ASE}, defined as the aggregate data rate per unit area.
We denote the \ac{ASE} as ${\rm ASE} (\tau) ~{\rm bits/s/Hz/m^2}$, which is given by~\cite{ZF-MU-MC}
\begin{align}\label{eq:ASE}
		{\rm ASE} (\tau) &=\frac{N_{\rm s}}{ \pi R_{\rm c}^2} \sum_{\kappa=1}^{N_{\rm a}} {\rm SE}_{\kappa}(\tau) ,
\end{align}
where $\frac{\pi R_{\rm c}^2}{N_{\rm s}}$ is the sector area.  

 \vspace{-2mm}
 
\section{General Framework of Performance Analysis}\label{sec:analy}

This section derives performance metrics defined in Sec.~\ref{subsec:metrics} for a multi-user single-cell network.  
We first present the spatial distributions of active users under the BPP. 
We then provide 
exact expressions for the CP, SE, and ASE, under the exact NF antenna pattern in \eqref{eq:antenna_pattern}.


\vspace{-2mm}

\subsection{Spatial Distributions of Active Users}\label{subsec:dist}

Based on the \ac{BPP} modeling of active users (referred to as nodes) in Sec.~\ref{subsec:network}, we provide the distributions of the distance from the BS at the origin to a random (non-ordered) node and to an ordered node, respectively, in the following.

\subsubsection{Non-Ordered Nodes}
Consider a random node in a \ac{BPP}, i.e.,  $\mathbf{u}=[\theta,r]^{\mathsf{T}} \in \Psi_{\rm a}$.  The angle $\theta$ is uniformly distributed within the sector, i.e., $\theta \sim \mathcal{U}[-\frac{\pi}{N_{\rm s}},\frac{\pi}{N_{\rm s}}]$.
The distance from the node to the BS at the origin satisfies $r \in [0,R_{\rm c}]$, where $R_{\rm c}$ is the cell radius. We denote the \ac{CDF} and \ac{PDF} of $r$ as $F_{R}(r)$ and $f_{R}(r)$, respectively. When $0\le r\le R_{\rm c}$, we have~\cite{BPP1}
 \begin{align}
		F_{R}(r) = \frac{r^2}{R_{\rm c}^2} \text{ and } f_{R}(r) =\frac{2r}{R_{\rm c}^2}. 
	\end{align}

\subsubsection{Ordered Nodes}\label{subsec:OrderedNode}

When nodes are ordered based on their distances to the origin, the angle remains uniformly distributed, while the distance distribution varies with the node order. 
Specifically, for the $\kappa$-th closest node located at $\mathbf{u}_{\kappa}=[\theta_\kappa,r_\kappa]^{\mathsf{T}}$, $\theta_\kappa \sim \mathcal{U}[-\frac{\pi}{N_{\rm s}},\frac{\pi}{N_{\rm s}}]$. The \ac{CDF} and \ac{PDF} of  $r_\kappa$, denoted by $F_{R_\kappa}(r_\kappa)$ and $f_{R_\kappa}(r_\kappa)$, are given by~\cite{BPP1,BPP2}
\begin{subequations}\label{eq:_Rk}
\begin{align}
\label{eq:CDF_Rk}
	F_{R_\kappa}(r_\kappa)
   &= 1-  \sum_{u=0}^{\kappa-1} {C}^{N_{\rm a}}_{u} \left( \frac{r_\kappa^2}{R_{\rm c}^2} \right)^{u}\left ( 1-\frac{r_\kappa^2}{R_{\rm c}^2} \right)^{N_{\rm a}-u}\!\!\!\!\!\!,  
	\\ \label{eq:PDF_Rk} f_{R_\kappa}(r_\kappa)
    &= \frac{2}{r_\kappa} \kappa {\ C}^{N_{\rm a}}_{\kappa} \left( \frac{r_\kappa}{R_{\rm c}} \right)^{2\kappa} \left (1-\left(\frac{r_\kappa}{R_{\rm c}}\right)^2\right)^{N_{\rm a}-\kappa}\!\!\!\!\!\!,
\end{align} 
\end{subequations}
for $0\le r_\kappa\le R_{\rm c}$.
When conditioning on the $\kappa$-th node being at distance $r_\kappa$, the remaining $N_{\rm a}-1$ nodes are no longer identically distributed~\cite{BPP1,BPP2}. Instead, they satisfy the ordering constraint: Exactly $\kappa-1$ nodes lie inside radius $r_\kappa$, referred to as {\em inner nodes}, and the remaining $N_{\rm a}-\kappa$ nodes lie outside radius $r_\kappa$, referred to as {\em outer nodes}. 
This is essential for analyzing the interference experienced by the $\kappa$-th active user at distance $r_\kappa$, which depends on the relative positions of inner and outer interferers.
To capture this constraint, we partition 
$\Psi_{\rm a}\setminus\{\mathbf{u}_\kappa\}$ into two subsets: $\Psi^{\kappa}_{\rm in}$ and $\Psi^{\kappa}_{\rm out}$, where
\begin{subequations}\label{eq:Psi_in_out}
 \begin{align}
   \Psi^{\kappa}_{\rm in} & = \{\mathbf{u}_{\kappa'}: \mathbf{u}_{\kappa'} \in \Psi_{\rm a}, \kappa'<\kappa\} = \{\mathbf{u}_{1}, ...,\mathbf{u}_{\kappa-1} \}, \\
   \Psi^{\kappa}_{\rm out} & =  \{\mathbf{u}_{\kappa'}: \mathbf{u}_{\kappa'} \in \Psi_{\rm a}, \kappa'>\kappa\} =\{\mathbf{u}_{\kappa+1}, ...,\mathbf{u}_{N_{\rm a}} \}.
\end{align}   
\end{subequations}
For nodes in $\Psi^{\kappa}_{\rm in} $ (or $\Psi^{\kappa}_{\rm out} $), their distributions are i.i.d.~\cite{BPP1}. 
Let $F_{R_{v}|r_{\kappa}}(r)$ and $f_{R_{v}|r_{\kappa}}(r)$, $v\in\{\rm in, out\}$ denote the conditional \ac{CDF} and PDF of the distance of a node in $\Psi^{\kappa}_{v} $, given that the $\kappa$-th node is at distance $r_\kappa$.
Specifically, for an inner node $\mathbf{u}= [\theta,r]^{\mathsf{T}} \in \Psi^{\kappa}_{\rm in}$, $\theta\sim\mathcal{U}[-\frac{\pi}{N_{\rm s}},\frac{\pi}{N_{\rm s}}]$ and for $ 0\le r< r_{\kappa}$,
\begin{align} \label{eq:Rin}
    F_{R_{\rm in}|r_{\kappa}}(r)=\frac{F_{R}(r)}{F_{R}(r_{\kappa})} \text{ and } 
	f_{R_{\rm in}|r_{\kappa}}(r)=\frac{f_{R}(r)}{F_{R}(r_{\kappa})}.
\end{align} 
For an outer node $\mathbf{u}= [\theta,r]^{\mathsf{T}} \in \Psi^{\kappa}_{\rm out}$, $\theta\sim\mathcal{U}[-\frac{\pi}{N_{\rm s}},\frac{\pi}{N_{\rm s}}]$ and for $r_{\kappa}\le r< R_{\rm c}$,
\begin{align}\label{eq:Rout}  
    \!\!F_{R_{\rm out}|r_{\kappa}}(r)\!=\!\frac{F_{R}(r)\!-\!F_{R}(r_{\kappa})}{1\!-\!F_{R}(r_{\kappa})} \text{ and }
	f_{R_{\rm out}|r_{\kappa}}(r)\!=\!\frac{f_{R}(r)}{1\!-\!F_{R}(r_{\kappa})}.\!
\end{align}  
\color{black}

\vspace{-6mm}

\subsection{Exact Expressions for CP, SE, and ASE}\label{subsec:cov}
Based on the above distance and angle distributions of the active users, we next derive the SIR-related performance metrics defined in \eqref{eq:condCP}-\eqref{eq:ASE}. 
 \begin{theorem} \label{them:cov}
Given the $\kappa$-th active user at $[\theta_\kappa, r_\kappa]^{\mathsf{T}}$, the conditional \ac{CP} in \eqref{eq:condCP} can be expressed as
\begin{align}\label{eq:condCP_exact}
		&{\rm CP}_{\kappa}(\tau |\theta_{\kappa},r_\kappa )=\frac{1}{2}- \frac{1}{2 {\rm j} \,\pi} \int_{0}^{\infty} \frac{1}{t} \, \big[ e^{-{\rm j}\, t \frac{1}{\tau}} \mathcal{L}_{\theta_{\kappa},r_\kappa}(-{\rm j}\, t) - 
		\nonumber\\&\qquad\qquad\qquad\quad e^{{\rm j}\, t \frac{1}{\tau}} \mathcal{L}_{\theta_{\kappa},r_\kappa}({\rm j}\, t)     \big] \mathrm{d}t.  
\end{align}
Here, $\mathcal{L}_{\theta_{\kappa},r_\kappa}(s)=\mathbb{E}_{\Psi_{\rm a}|\theta_\kappa, r_\kappa}\left\{\exp{\left( -s  I_{\kappa}  \right) } \right\}$ is the Laplace transform of the normalized aggregate interference $I_{\kappa}$, which is given by
\begin{align}\label{eq:Laplace}
\begin{split}
   \mathcal{L}_{\theta_{\kappa},r_\kappa}(s)&= (\mathcal{L}_{\theta_{\kappa},r_\kappa}^{\rm in}(s))^{\kappa-1} (\mathcal{L}_{\theta_{\kappa},r_\kappa}^{\rm out}(s))^{N_{\rm a}-\kappa} ,
\end{split} 
\end{align}
where  
\begin{align} \nonumber 
\begin{split}
\mathcal{L}_{\theta_{\kappa},r_\kappa}^{v}(s)  
		=  \!\int_{-\frac{\pi}{N_{\rm s}}}^{\frac{\pi}{N_{\rm s}}} \!\int_{0}^{R_{\rm c}} \!\!\! {\rm e}^{ -s \mathcal{G} (\theta,r;\theta_\kappa,r_{\kappa}) } f_{R_{v}|r_{\kappa}}(r) \frac{N_{\rm s}}{2\pi} \mathrm{d} r \mathrm{d} \theta,  
\end{split}
\end{align}
where $v\in \{\rm in, out \}$, $f_{R_{v}|r_{\kappa}}(r)$ is given in \eqref{eq:Rin}-\eqref{eq:Rout}, and 
$\mathcal{G} (\cdot)$ is the exact antenna pattern in \eqref{eq:antenna_pattern}.
The overall \ac{CP} in \eqref{eq:overCP} is
 \begin{align}\label{eq:CP_exact}
	\begin{split}
&{\rm CP}_{\kappa} (\tau)
        =\! \int_{-\frac{\pi}{N_{\rm s}}}^{\frac{\pi}{N_{\rm s}}} \!\int_{0}^{R_{\rm c}} \!\!\!\!{\rm CP}_{\kappa}(\tau |\theta_{\kappa},r_\kappa )  f_{R_{\kappa}}(r_\kappa) \frac{N_{\rm s}}{2\pi} \mathrm{d} r_\kappa \mathrm{d} \theta_{\kappa},
	\end{split}
\end{align}
where $f_{R_{\kappa}}(r_\kappa)$ is given in \eqref{eq:PDF_Rk}. 
	\begin{IEEEproof}
		See Appendix~\ref{app:them:cov}.
	\end{IEEEproof}
\end{theorem}

Substituting \eqref{eq:condCP_exact} into \eqref{eq:condAR} yields the conditional \ac{SE}. Substituting \eqref{eq:CP_exact} into \eqref{eq:overAR} and \eqref{eq:ASE} yields the overall \ac{SE} and the \ac{ASE}, respectively.  Note that Theorem~\ref{them:cov} provides exact expressions for performance metrics by using the exact NF antenna pattern in \eqref{eq:antenna_pattern}. 
However, this complicated pattern leads to an intractable expression for the Laplace transform of the interference in \eqref{eq:Laplace}. As a result, the CP, SE, and ASE involve multiple nested integrals over angle and distance, making the \ac{NF} performance analysis highly complicated and computationally demanding.
\textcolor{black}{Moreover, the lack of analytical tractability makes it difficult to extract explicit system-level insights, such as performance scaling laws with respect to key parameters.}
To address this challenge, we propose an approximate \ac{NF} antenna pattern in the next section.

\vspace{-2mm}

\section{Approximate Antenna Pattern}\label{sec:approx_pattern}

The exact NF antenna pattern in \eqref{eq:antenna_pattern} varies continuously with both angle and distance, resulting in a highly intricate antenna gain distribution that complicates analytical performance evaluation.
However, Fig.~\ref{fig:pattern} reveals a key property of NF beamfocusing: Most of the radiated energy is concentrated around the focal point within a finite beam depth (distance range) and beam width (angular range), while the gain outside this region is relatively small. 
\color{black}
As discussed in Sec.~\ref{subsec:ActPattern}, the antenna pattern relates the interference at the tagged user to the spatial locations of interferers. 
Consequently, due to the beamfocusing property, the dominant interference primarily arises from interferers located within a limited region around the tagged user.
Motivated by this observation, we approximate the exact continuous pattern using a discretized representation, termed the near-field multi-level antenna pattern (NF-MLAP). It quantizes the gain into a finite number of distance-domain and angular-domain levels to provide a tractable representation that captures the dominant interference characteristics.
\color{black}

To construct the NF-MLAP, it is essential to first understand how the antenna gain varies along the {\em radial direction (distance leakage)} and across {\em angles (angular leakage)} around the focal point.
In what follows, we first examine how the gain evolves as the observation point moves away from the focal point in distance while keeping the angle fixed. We then characterize the angular leakage independently of distance effects. These analyses guide the construction of distance- and angular-domain discretization regions and their corresponding gain levels, which enable the formulation of the approximate polar-domain antenna pattern.
\textcolor{black}{Moreover, we compare the approximate and exact patterns via numerical results to demonstrate that the proposed approximation captures the key characteristics of the exact antenna pattern.} The default array configuration is the same as that in Fig.~\ref{fig:pattern}, i.e., the ULA along the $\rm y$-axis with its center at the origin.
\color{black}

\vspace{-2mm}

\subsection{Distance-Domain Antenna Pattern} \label{subsec:BD} 

Given a focal point $[\theta_\kappa,r_\kappa]^\mathsf{T}$, to capture the radial leakage of beam energy, we investigate how the beam gain decreases as the observation point $[\theta,r]^\mathsf{T}$ varies in distance while maintaining the same angle, i.e., $\theta = \theta_\kappa$. 
The resulting {\em distance-domain antenna pattern}, denoted by $\mathcal{G}_{\rm D}(\beta_{\theta_\kappa,r_\kappa,r})  \triangleq   \mathcal{G} (\theta_\kappa,r;\theta_\kappa,r_\kappa)$, isolates the gain degradation induced purely by distance mismatch. It can be expressed as \cite[Lemma 1]{cui2022channel}
\begin{align}\label{eq:GR_exact}
		&  \mathcal{G}_{\rm D}(\beta_{\theta_\kappa,r_\kappa,r}) 
         = \bigg|\frac{C(\beta_{\theta_\kappa,r_\kappa,r}) +{\rm j} S(\beta_{\theta_\kappa,r_\kappa,r})} {\beta_{\theta_\kappa,r_\kappa,r}}\Bigg|^2,
\end{align}
where $\beta_{\theta_\kappa,r_\kappa,r}=\sqrt{\frac{N^2d^2(1-\sin^2\theta_\kappa)}{2\lambda} \big|\frac{1}{r_\kappa}-\frac{1}{r}\big|}$, and
$C(\beta)=\int_{0}^{\beta}\cos(\frac{\pi}{2}t^2) \mathrm{d}t$ and $S(\beta)=\int_{0}^{\beta}\sin(\frac{\pi}{2}t^2) \mathrm{d}t$ are the Fresnel functions.
Define $\mathcal{D}_{\theta_\kappa,r_\kappa}=(D_{\theta_\kappa,r_\kappa}^{\rm left},D_{\theta_\kappa,r_\kappa}^{\rm right})$ as the distance range where the gain remains within $\gamma$ [dB] of the peak, i.e., 
\begin{align}\label{eq:GR-BD}
    \mathcal{G}_{\rm D}(\beta_{\theta_\kappa,r_\kappa,r})>10^{\frac{\gamma}{10}}, \text{ for } r\in \mathcal{D}_{\theta_\kappa,r_\kappa}.
\end{align}
$\mathcal{D}_{\theta_\kappa,r_\kappa}$ corresponds to the beam depth interval, whose boundaries are given by~\cite{BWBD} 
\begin{align}\label{eq:Dleft-Dright}
    D_{\theta_\kappa,r_\kappa}^{\rm left}&=\frac{ r_\kappa  D_{\theta_\kappa}}{D_{\theta_\kappa}+r_\kappa},
D_{\theta_\kappa,r_\kappa}^{\rm right}=\begin{cases}
\frac{ r_\kappa  D_{\theta_\kappa}}{D_{\theta_\kappa}-r_\kappa}, & \!\!r_\kappa< D_{\theta_\kappa},\\
\infty, & \!\!\text{otherwise}.
\end{cases}
\end{align}
Here, $D_{\theta_\kappa}=\frac{ N^2 d^2 (1-\sin^2\theta_\kappa) }{2\lambda{\beta}_{\gamma}}$, where $\beta_\gamma$ is a constant for a given $\gamma$, obtained by numerically solving $\mathcal{G}_{\rm D}(\beta_\gamma)=10^{\frac{\gamma}{10}}$~\cite{cui2022channel}, e.g., $\beta_\gamma=1.3$ corresponds to $\gamma\approx-3~\rm dB$.
The corresponding beam depth is 
\begin{align}\label{eq:BD}
    {\rm BD}(\theta_\kappa,r_\kappa)=\!D_{\theta_\kappa,r_\kappa}^{\rm right}\!-\!D_{\theta_\kappa,r_\kappa}^{\rm left}\!\!=\!\begin{cases}
\frac{2r_\kappa^2 D_{\theta_\kappa}}{D_{\theta_\kappa}^2-r_\kappa^2}, & \!\!r_\kappa < D_{\theta_\kappa},\\
\infty, & \!\! r_\kappa \ge D_{\theta_\kappa}.
\end{cases}
\end{align}

From \eqref{eq:Dleft-Dright}, $D_{\theta_\kappa}$ is proportional to $N^2$; when $r_\kappa < D_{\theta_\kappa}$, 
the beam depth is finite; when $r_\kappa\ge D_{\theta_\kappa}$, the beam depth is infinite. Therefore, reducing the antenna number ($N$) or increasing the focal distance ($r_\kappa$) leads to a gradual expansion of the beam depth.  Eventually, the beam depth reaches infinity, which signifies the transition to the \ac{FF} scenario where the beam lacks distance dependency.

\begin{figure*}[t!]
\vspace{-3mm}
\begin{minipage}{.32\textwidth}
    \centering
     \subfloat[Varying focal distance $r_\kappa$.]{\includegraphics[width=1.1\columnwidth]{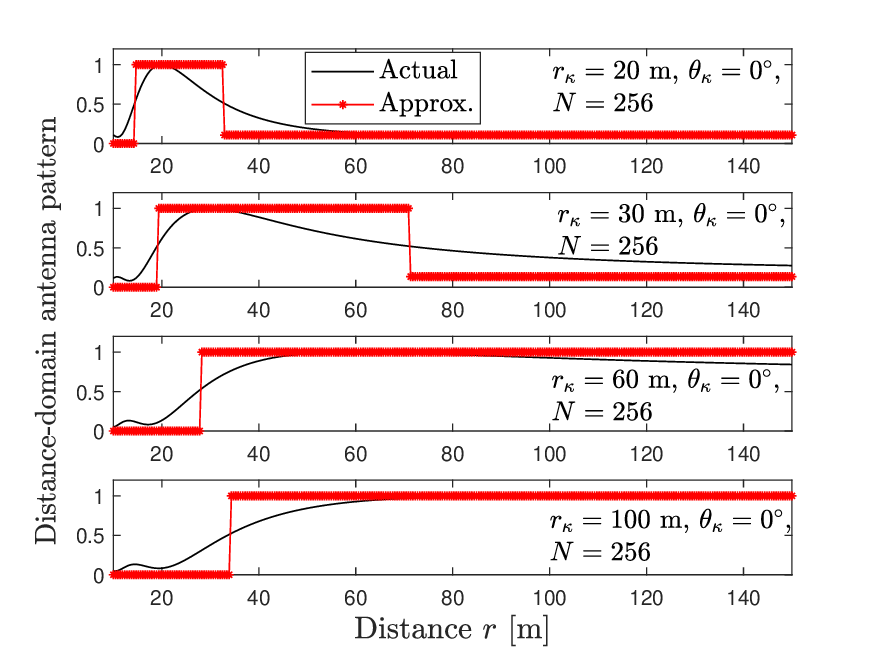}\label{fig:N256range}}
     
\vspace{-3mm}
 \subfloat[Varying antenna number $N$.]{\includegraphics[width=1.1\columnwidth]{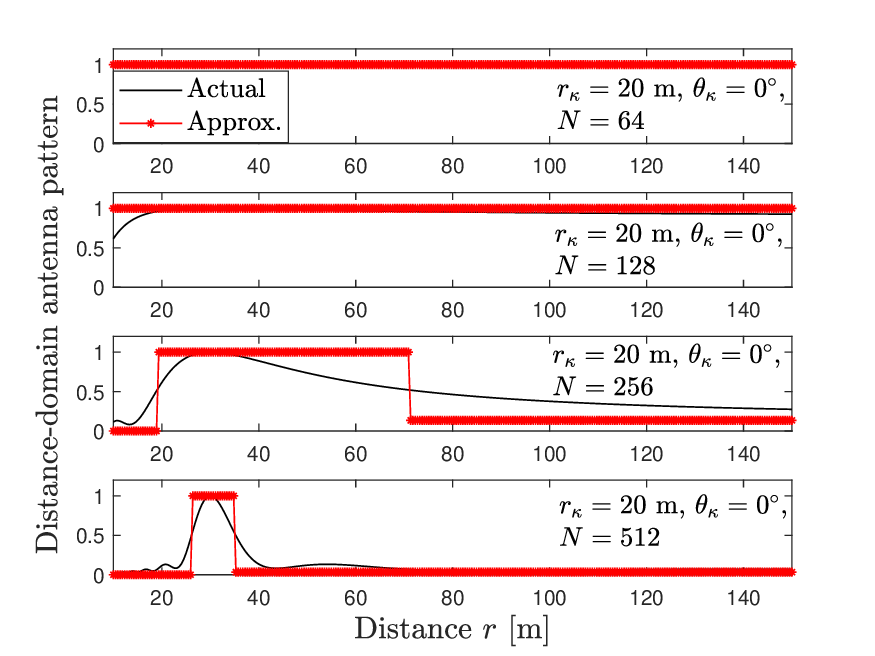}\label{fig:Nrange}}
  \vspace{-1mm}
 \caption{Actual and approximate distance-domain antenna patterns, i.e., $\mathcal{G}_{\rm D}(\beta_{\theta_\kappa,r_\kappa,r})$ in \eqref{eq:GR_exact} and $\widetilde{\mathcal{G}}_{\rm D}(\beta_{\theta_\kappa,r_\kappa,r})$ in \eqref{eq:GRapprox}.} 
 \label{fig:range} 
\end{minipage}
\hfill
\begin{minipage}{.32\textwidth}
    \centering
 \subfloat[Varying steering direction $\theta_\kappa$.]{\includegraphics[width=1.1\columnwidth]{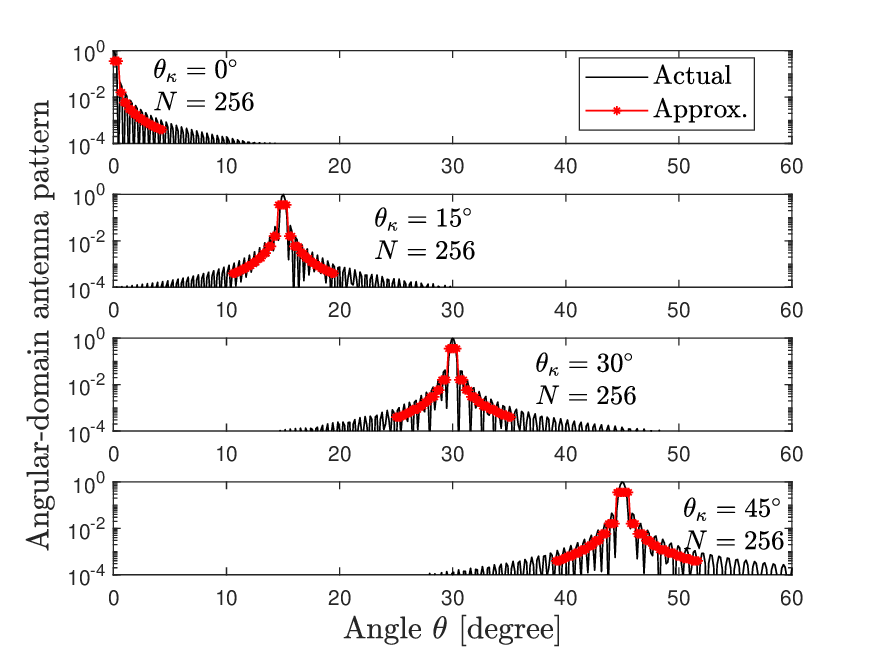}\label{fig:N256angle}}

\vspace{-3mm}
 \subfloat[Varying antenna number $N$.]{\includegraphics[width=1.1\columnwidth]{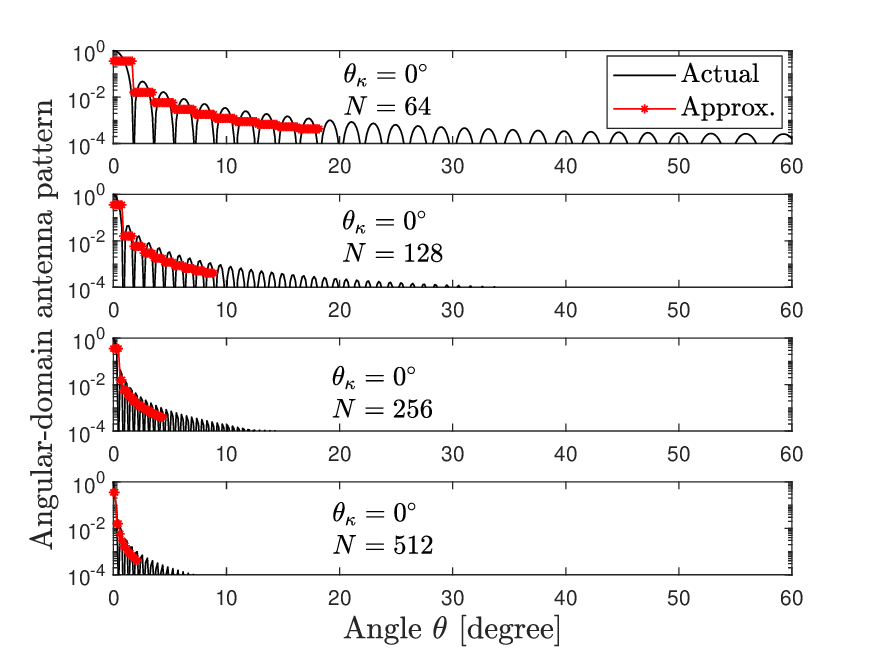}\label{fig:Nangle}}
 \vspace{-1mm}
 \caption{Actual and approximate angular-domain antenna patterns, i.e., $\mathcal{G}_{\rm A}(\theta;\theta_\kappa)$ in \eqref{eq:angular} and $\widetilde{\mathcal{G}}_{{\rm A},M}(\theta;\theta_\kappa)$ in \eqref{eq:GAapprox} with $M=10$.}%
 \label{fig:angle} 
    \end{minipage}
\hfill
\begin{minipage}{.32\textwidth}
  \vspace{3.5mm}
		\centering
 \subfloat[Varying focal distance $r_\kappa$.]{\includegraphics[width=1.1\columnwidth]{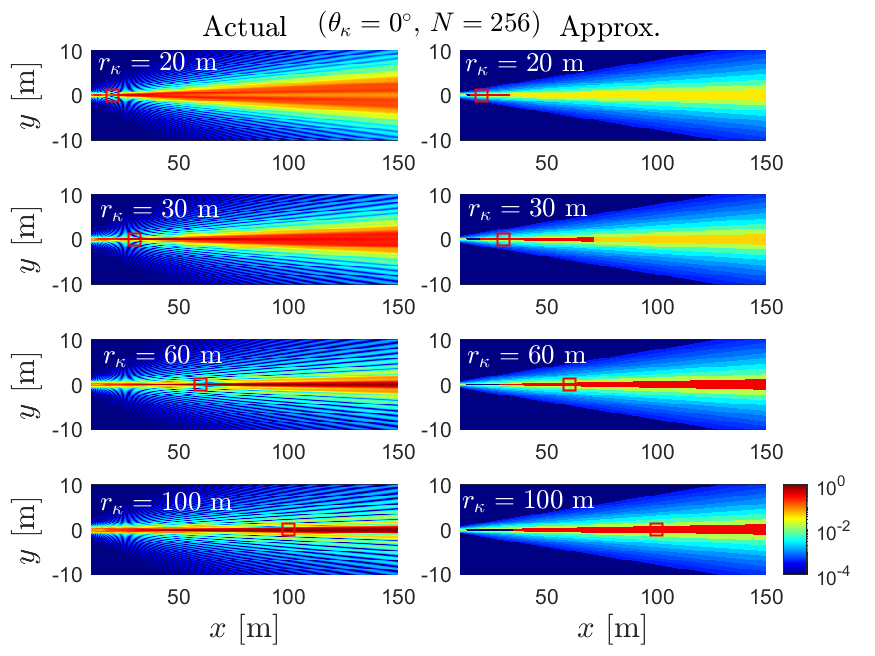}\label{fig:N256polar_log}}
 
\vspace{-3mm}
 \subfloat[Varying antenna number $N$.]{\includegraphics[width=1.1\columnwidth]{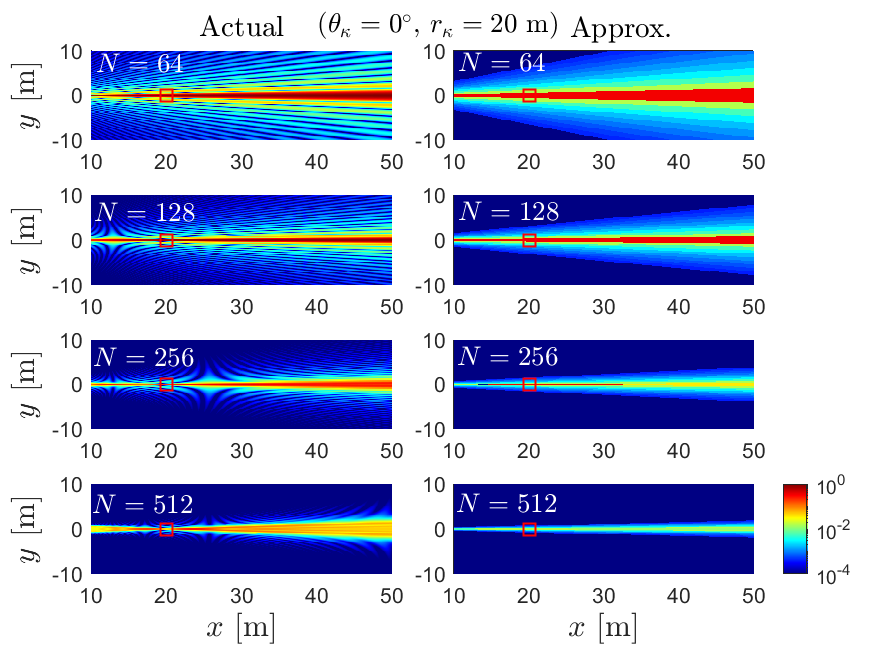}\label{fig:Npolar_log}}
 \vspace{-1mm}
 \caption{Actual and approximate polar-domain antenna patterns, i.e., $\mathcal{G} (\theta, r ; \theta_\kappa, r_\kappa)$ in \eqref{eq:antenna_pattern} and $\widetilde{\mathcal{G}}_{M} (\theta, r ; \theta_\kappa, r_\kappa)$ in \eqref{eq:NF_MLAP}, where the red squares mark the focal points and $M=10$.}
 \label{fig:polar} 
    \end{minipage}
    \vspace{-5mm}
\end{figure*}

\begin{approximation}\label{approx:dist}
The distance-domain pattern $\mathcal{G}_{\rm D}(\beta_{\theta_\kappa,r_\kappa,r})$ exhibits three characteristic behaviors in Fig.~\ref{fig:range}. 
First, within $\mathcal{D}_{\theta_\kappa,r_\kappa}=(D_{\theta_\kappa,r_\kappa}^{\rm left},D_{\theta_\kappa,r_\kappa}^{\rm right})$, the beam gain remains close to its maximum. Second, for $r>D_{\theta_\kappa,r_\kappa}^{\rm right}$, the gain gradually approaches a non-zero asymptotic value. As $r\to\infty$, $\beta_{\theta_\kappa,r_\kappa,r}$ approaches $ \sqrt{\frac{N^2d^2\cos^2\theta_\kappa}{2\lambda} \frac{1}{r_\kappa}}$ and thus $\mathcal{G}_{\rm D}(\beta_{\theta_\kappa,r_\kappa,r})$ converges to $g_{\theta_\kappa,r_\kappa}\triangleq\mathcal{G}_{\rm D}\left(\sqrt{\frac{N^2d^2\cos^2\theta_\kappa}{2\lambda} \frac{1}{r_\kappa}}\right)$. 
Third, for $r<D_{\theta_\kappa,r_\kappa}^{\rm left}$, the gain rapidly drops and becomes negligible.  
Based on these observations, we quantize $\mathcal{G}_{\rm D}(\beta_{\theta_\kappa,r_\kappa,r})$ into three levels as
\begin{align}\label{eq:GRapprox}
	\begin{split}
        \widetilde{\mathcal{G}}_{\rm D}(\beta_{\theta_\kappa,r_\kappa,r})
        \! = \!\!\left\{\begin{matrix}
		1, \!\!&\!\! {\rm {if}}\,  r\!\in\! (D_{\theta_\kappa,r_\kappa}^{\rm left},D_{\theta_\kappa,r_\kappa}^{\rm right}),\hfill\\
		g_{\theta_\kappa,r_\kappa}, \!\!& \!\!  {\rm {if}}\, r\!\in\! [D_{\theta_\kappa,r_\kappa}^{\rm right},\infty),
		\hfill\\
		0,\!\! &\!\! {\rm otherwise.}\hfill
	\end{matrix}\right.
	\end{split}
\end{align}    
\end{approximation}

Fig.~\ref{fig:range} compares the actual and approximate distance-domain antenna patterns, i.e., $ \mathcal{G}_{\rm D}(\cdot)$ and $ \widetilde{\mathcal{G}}_{\rm D}(\cdot)$, for a beam focused on $[\theta_\kappa,r_\kappa]^{\mathsf{T}}$ with different values of $r_\kappa$ and $N$.
We see that the approximate pattern captures the high-gain region through the beam depth. Moreover, as $N$ decreases or $r_\kappa$ increases, the beam depth gradually widens and eventually becomes infinite, consistent with the NF-to-FF transition shown in \eqref{eq:Dleft-Dright}-\eqref{eq:BD}.

\vspace{-2mm}

\subsection{Angular-Domain Antenna Pattern} \label{subsec:BW}
 
To characterize how the antenna gain drops as the angle $\theta$ deviates from $\theta_\kappa$, i.e., angular leakage of beam energy, we isolate the angular component $\Phi^{\rm ang}_n$ of \eqref{eq:antenna_pattern}. Specifically, when $\frac{1-\sin^2\theta_\kappa}{2r_\kappa}=\frac{1-\sin^2\theta}{2r}$, the distance-dependent term $\Phi^{\rm dist}_n$ of ~\eqref{eq:antenna_pattern} cancels out, leaving a gain expression that depends only on the angle deviation.
The corresponding pattern, referred to as {\em angular-domain antenna pattern}, is the same as the \ac{FF} antenna pattern in~\eqref{eq:FF}, which is given by~\cite{cui2022channel,LDMA}
\begin{align}\label{eq:angular}
		\mathcal{G}_{\rm A}(\theta;\theta_\kappa) 
        =\frac{\sin^2(\pi N \phi)}{N^2 \sin^2(\pi\phi)} \triangleq \mathcal{G}_{\rm A}(\phi),
\end{align}
where $\phi\triangleq \frac{1}{2}(\sin\theta-\sin\theta_\kappa)$. We define $\vartheta = \frac{1}{2}\sin \theta$ (or $\vartheta_\kappa = \frac{1}{2}\sin \theta_\kappa$) as the {\em spatial angle} corresponding to the physical angle $\theta$ (or $\theta_\kappa$)~\cite{ActualAntenna}. Thus, $\phi = \vartheta - \vartheta_\kappa$ represents the spatial angle deviation from the focused direction.
Note that this pattern is a squared Dirichlet kernel normalized by $N^2$, which has several crucial parameters, e.g., the beam width, the maxima of each sidelobe, and the nulls~\cite{AntennaTheory,ActualAntenna}. 
In particular, the first nulls of \eqref{eq:angular} occur at $\phi=\pm \frac{1}{ N}$, which implies that the first-null beam width (i.e., the width of the mainlobe measured between the first nulls) is $\frac{2}{N}$.

Several approximations of $\mathcal{G}_{\rm A}(\cdot)$ have been studied in the literature.
The flat-top pattern~\cite{FlatTop} provides a binary quantization based on the beam width. The multi-constant discrete pattern~\cite{DiscreteAntenna,chen2024joint} offers a more precise approximation by quantizing $\mathcal{G}_{\rm A}(\cdot)$ into $ \left \lfloor  \frac{N}{2} \right \rfloor $ levels, which covers the mainlobe and all sidelobes, with the $i$-th level corresponding to the scaled maximum of the $i$-th lobe, $i=1,..., \left \lfloor  \frac{N}{2} \right \rfloor $.


\begin{approximation}\label{approx:angle}
Since sidelobe levels decay rapidly, it is sufficient to approximate only the mainlobe and a limited number of sidelobes of the angular-domain pattern $\mathcal{G}_{\rm A}(\theta;\theta_\kappa)= \mathcal{G}_{\rm A}(\phi)$. 
Therefore, we quantize $\mathcal{G}_{\rm A}(\phi)$ into $M$ levels as
 \begin{align}\label{eq:GAapprox}
    \begin{split}
		\widetilde{\mathcal{G}}_{{\rm A},M}(\theta;\theta_\kappa) \!=\!\widetilde{\mathcal{G}}_{{\rm A},M} (\phi) 
      \! =\! \begin{cases} 
			g_1,  & \!\!\!\text{ if } |\phi |\le \frac{1}{N}, \\
			g_m, & \!\!\!\text{ if } \frac{m-1}{N}<|\phi |\le \frac{m}{N},\\
			0,  & \!\!\!\text{ otherwise},
		\end{cases}        
    \end{split}
	\end{align}
where $M \le \left \lfloor  \frac{N}{2} \right \rfloor $, $\phi=\frac{1}{2}(\sin\theta-\sin\theta_\kappa)$,
$g_m=\frac{\delta}{2}\mathcal{G}_{\rm A}(\frac{2m-1}{2N})$, for $m\in\{2,3,..., M\}$, $g_1=\frac{\delta}{2}\mathcal{G}_{\rm A}(0)$, and $\delta=\frac{1}{\sqrt{2}}$ compensates for roll-off characteristics in \eqref{eq:angular}. 
\end{approximation}


Fig.~\ref{fig:angle} presents the actual and approximate angular-domain antenna patterns, i.e., $\mathcal{G}_{\rm A} (\theta;\theta_\kappa)$ and $\widetilde{\mathcal{G}}_{{\rm A},M} (\theta;\theta_\kappa)$ with $M=10$, for different $\theta_\kappa$ and $N$. 
We see that as $N$ increases, the beam width significantly decreases. Moreover, sidelobe levels decay rapidly, and the first $10$ lobes provide a sufficiently accurate representation of the angular-domain antenna pattern.

\vspace{-2mm}

\subsection{Near-Field Multi-Level Antenna Pattern (NF-MLAP)}\label{subsec:NF-MLAP}
\color{black} 
The above distance- and angular-domain analyses separately characterize the energy leakage along each dimension. 
However, the exact NF antenna pattern in \eqref{eq:antenna_pattern} jointly depends on the distance mismatch and the angle deviation.
Specifically, it is determined by the coherent summation of the distance mismatch-induced quadratic phase component $\Phi_n^{\rm dist}$ and the angle deviation-induced linear phase component $\Phi_n^{\rm ang}$ across the array aperture.
To construct a tractable polar-domain approximation, it is therefore essential to understand how the two components contribute to phase variation across the array.


From~\eqref{eq:antenna_pattern}, $\Phi_n^{\rm dist}=\frac{2\pi}{\lambda}(nd)^2(\frac{1-\sin^2\theta_\kappa}{2r_\kappa}-\frac{1-\sin^2\theta}{2r})$ satisfies 
\begin{align}
    |\Phi_n^{\rm dist}|&
    \overset{(a)}{\le}\frac{2\pi}{\lambda}(nd)^2\left(\left|\frac{1-\sin^2\theta_\kappa}{2r_\kappa}\right|+\left|\frac{1-\sin^2\theta}{2r}\right|\right)
     \nonumber\\& \overset{(b)}{\le} \frac{2\pi}{\lambda}(Nd)^2\left(\frac{1}{2r_\kappa}+ \frac{1}{2r}\right)
    \overset{(c)}{=}\mathcal{O}\left(\frac{N^2}{r'}\right), 
\end{align}
where (a) is from $n\le N$ and $|a_1-a_2|\le |a_1|+|a_2|$ for $a_1,a_2\in\mathbb{R}$ and $a_1,a_2\ge 0$, (b) is from $n\le N$ and $|1-\sin^2\theta|\le1$, and (c) is from $r'\triangleq\min\{r_\kappa,r\}$. 
For users located in the radiative NF region, $r'$ typically lies between the Fresnel distance $0.62\sqrt{\tfrac{D^3}{\lambda}}$ and the Rayleigh distance $\tfrac{2D^2}{\lambda}$, where $D=(N-1)d$ is the array aperture~\cite{NFtutorial}. 
Hence, $r'$ scales from $\mathcal{O}(N^\frac{3}{2})$ to $\mathcal{O}(N^2)$, which implies that the variation of the distance-dependent phase term across the array satisfies $|\Phi_n^{\rm dist}|=\mathcal{O}(1) \text{ to }\mathcal{O}(N^\frac{1}{2})$. 

From~\eqref{eq:antenna_pattern}, $\Phi_n^{\rm ang}=\frac{4\pi}{\lambda}n d\phi$ satisfies
\begin{align}
     |\Phi_n^{\rm ang}|\le \frac{4\pi}{\lambda}N d \left|\phi\right|=\mathcal{O}(N|\phi|).
\end{align}
The behavior of $|\Phi_n^{\rm ang}|$ depends on whether the angular mismatch $|\phi|$ lies within or outside the mainlobe.
{\em (i)} In the {\em mainlobe region} with $|\phi|<\frac{1}{N}$, $N|\phi|<1$, implying $|\Phi_n^{\rm ang}|=\mathcal{O}(1)$, so the angular phase varies slowly across the array. Since the distance-dependent phase remains $\mathcal{O}(1)$ to $\mathcal{O}(N^{\frac{1}{2}})$, it is generally non-negligible relative to the angular component.  As a result, both components contribute to the phase alignment across antennas and jointly determine the coherent summation. {\em (ii)} In the {\em sidelobe region} with $|\phi|>\frac{1}{N}$, the angular phase enters a growing regime $\mathcal{O}(N|\phi|)$, which can reach up to $\mathcal{O}(N)$, leading to significant phase dispersion across the array.
In contrast, the distance-dependent phase varies at most on the order of $\mathcal{O}(N^{\frac{1}{2}})$, and thus constitutes a lower-order perturbation. 
Consequently, as the angular mismatch moves away from the mainlobe boundary,  the angular term progressively dominates the phase dispersion, while the contribution of the distance term becomes negligible in comparison.


These observations motivate a region-dependent approximation of the NF antenna pattern.   
Specifically, in the mainlobe region, both angular and distance dimensions are jointly discretized to capture the finite beam width and beam depth.
In contrast, in the sidelobe region, the antenna pattern is primarily governed by angular variations and can be approximated using angular-domain multi-level gains.
Note that this approximation does not imply that the sidelobe gain is strictly independent of distance.
Since the sidelobe gain is inherently weak, incorporating additional distance-domain refinements yields limited analytical benefit. 
\textcolor{black}{Therefore, this construction preserves the dominant polar-domain structure of the exact \ac{NF} beam pattern, as formulated below.}
\color{black}


\color{black} \begin{prop}\label{Approx:NF-MLAP}
Based on the NF beamfocusing property with finite beam depth and beam width, we approximate the exact NF antenna pattern in \eqref{eq:antenna_pattern} by the following NF-MLAP as
\begin{align}\label{eq:NF_MLAP}
\widetilde{\mathcal{G}}_{M}(\theta,r;\theta_\kappa,r_\kappa)= 
\begin{cases} 
g_0,  & \text{if } (\theta,r) \in \mathcal{R}_0 (\theta_\kappa,r_\kappa), \\
g_1,  & \text{if } (\theta,r) \in \mathcal{R}_1 (\theta_\kappa,r_\kappa), \\
g_m,  & \text{if } (\theta,r) \in \mathcal{R}_m (\theta_\kappa,r_\kappa),\\
g_{M+1},  & \text{otherwise}.
\end{cases}
\end{align}
Here, $2\le m\le M$, $M\le \left\lfloor \frac{N}{2}\right\rfloor$, $g_0\triangleq g_1 g_{\theta_\kappa,r_\kappa}$, $g_{\theta_\kappa,r_\kappa}$ is defined in Sec.~\ref{subsec:BD},
$g_m$ is defined in \eqref{eq:GAapprox}, $g_{M+1}=0$,
 and 
\begin{align}
\mathcal{R}_0 (\theta_\kappa,r_\kappa)
&\triangleq
\left\{(\theta,r): |\phi|\le \frac{1}{N},~ r \in [D_{\theta_\kappa,r_\kappa}^{\rm right},\infty)\right\},
\nonumber\\
\mathcal{R}_1 (\theta_\kappa,r_\kappa)
&\triangleq
\left\{(\theta,r): |\phi|\le \frac{1}{N},~ r \in (D_{\theta_\kappa,r_\kappa}^{\rm left},D_{\theta_\kappa,r_\kappa}^{\rm right})\right\},
\nonumber\\
\mathcal{R}_m (\theta_\kappa,r_\kappa)
&\triangleq
\left\{(\theta,r): \frac{m-1}{N}<|\phi|\le \frac{m}{N}\right\},
\nonumber
\end{align}
where $2\le m\le M$, $\phi=\frac{1}{2}\sin\theta-\frac{1}{2}\sin\theta_\kappa$, 
and $D_{\theta_\kappa,r_\kappa}^{\rm left}$ and $D_{\theta_\kappa,r_\kappa}^{\rm right}$ are defined in Sec.~\ref{subsec:BD}.
\end{prop}

Under the NF-MLAP, the spatial domain is partitioned into several regions, each associated with a specific interference gain level.
Specifically, 
$\mathcal{R}_i(\theta_\kappa,r_\kappa)$ represents the $i$-th \emph{interfering region},  
i.e., if an interferer lies within this region, the associated beam yields the approximate interference gain $g_i$ toward the tagged user. 
In particular, $\mathcal{R}_1(\theta_\kappa,r_\kappa)$ corresponds to the \emph{\ac{NF} beamfocusing region} jointly characterized by the beam width and beam
depth, where interferers are close to the tagged user in both angle and distance, 
thus producing the strongest interference gain. 
The region $\mathcal{R}_0(\theta_\kappa,r_\kappa)$ includes interferers that are angularly close to the tagged user but sufficiently far in distance, 
resulting in a reduced interference gain. 
The regions $\{\mathcal{R}_m(\theta_\kappa,r_\kappa):2\le m\le M\}$ correspond to the \emph{sidelobe regions}, where the angular mismatch exceeds the beam width, and the interference gain rapidly decreases with $m$. 
Note that the parameter $M$ controls the number of quantization levels used to approximate the sidelobe structure, and its selection for accurate performance evaluation will be discussed in Sec.~\ref{subsec:M}.

\color{black}

Fig.~\ref{fig:polar} presents the actual and approximate polar-domain
antenna patterns, i.e., ${\mathcal{G}}(\cdot)$ and $\widetilde{\mathcal{G}}_{M}(\cdot)$ with $M=10$. We see that the NF-MLAP preserves the NF beamfocusing property by jointly modeling beam depth and width.
Moreover, as discussed in Sec.~\ref{subsec:BD}, decreasing the antenna number ($N$) or increasing the focal distance ($r_\kappa$) enlarges the beam depth, and when it becomes infinite, the distance dependence vanishes and the NF-MLAP $\widetilde{\mathcal{G}}_{M}(\cdot)$  naturally reduces to an angular-only pattern $\widetilde{\mathcal{G}}_{{\rm A},M} (\cdot)$.
This suggests that the NF-MLAP inherently preserves the NF-to-FF transition.

\vspace{-1mm}

\section{Performance Analysis Under Proposed Pattern}\label{sec:approx_analy}
 
In this section, we provide approximate expressions for the performance metrics using the proposed NF-MLAP. We then derive upper bounds to further reduce the computational complexity. 
Moreover, we extend the SIR-related analysis to the SINR-related analysis.

\vspace{-2mm}

\subsection{Performance Analysis Under NF-MLAP}

With the NF-MLAP $\widetilde{\mathcal{G}}_{M}(\theta,r;\theta_\kappa,r_\kappa)$ in \eqref{eq:NF_MLAP}, for a given focal point $[\theta_\kappa, r_\kappa]^{\mathsf{T}}$, 
the normalized beamforming gain is approximated as a constant within each interval of $\phi$ and $r$. These constants form the set $\{g_i:i\in\{0,1,...,M+1\}\}$.
For performance analysis, we begin by providing the probability that the normalized gain takes the value $g_i$ under the NF-MLAP based on the distributions of $\phi$ and $r$. 
The distribution of $r$ has been provided in Sec.~\ref{subsec:dist}. 
To obtain the distribution of $\phi=\vartheta - \vartheta_\kappa$, we first provide the \ac{CDF} and \ac{PDF} of the spatial angle $\vartheta=\frac{1}{2}\sin \theta$, denoted as $F_\Theta(\vartheta)$ and $f_\Theta(\vartheta)$, respectively. 
Based on the uniformly distributed physical angle $\theta\sim\mathcal{U}[-\frac{\pi}{N_{\rm s}},\frac{\pi}{N_{\rm s}}]$, when $-\frac{1}{2}\sin\frac{\pi}{N_{\rm s}}\leq \vartheta\leq\frac{1}{2}\sin\frac{\pi}{N_{\rm s}}$,  
\begin{align} \label{eq:sptial_angle}
    \!\!F_\Theta(\vartheta)\!=\! \frac{N_{\rm s}}{2\pi} \!\left[\!\arcsin(2\vartheta)\!+\!\frac{\pi}{N_{\rm s}}\!\right] \text{and }
    f_\Theta(\vartheta)\!= \!\frac{N_{\rm s}}{\pi\sqrt{1 \!-\! 4 \vartheta ^{2}}}.\!
\end{align}
Conditioned on $\vartheta_\kappa$, the \ac{CDF} and \ac{PDF} of $\phi$ are given by 
\begin{align}\label{cdf:pdf:phi}
    F_{\Phi|\vartheta_\kappa}(\phi)=F_\Theta(\phi+\vartheta_\kappa) \text{ and }f_{\Phi|\vartheta_\kappa}(\phi)=f_\Theta(\phi+\vartheta_\kappa).
\end{align}

\begin{lemma}\label{lemma:prob_Gain}
With the NF-MLAP in \eqref{eq:NF_MLAP}, \textcolor{black}{given the tagged user located at $[\theta_\kappa, r_\kappa]^{\mathsf{T}}$, 
the probability that an inner (or outer) interferer lies in the region $\mathcal{R}_i (\theta_\kappa,r_\kappa)$, thereby yielding the interference gain $g_i$, is denoted as $p_i^{\rm in}$ (or $p_i^{\rm out}$), where $i\in\{0,1,..., M+1\}$.}
Specifically, $p_0^{v}$ and $p_1^{v}$, $v\in \{\rm in,out\}$, depend on both the angle and distance distributions, i.e., 
\begin{align}\label{eq:p0}
p_0^{v} & =  F_{\Phi|\vartheta_\kappa}(\phi)\big|_{-\frac{1}{N}}^{\frac{1}{N}}\times F_{R_{v}|r_{\kappa}}(r)\big|_{D_{r_\kappa,\theta_\kappa}^{\rm right}}^{\infty},
\end{align}  
\begin{align}\label{eq:p1}
p_1^{v} &=F_{\Phi|\vartheta_\kappa}(\phi)\big|_{-\frac{1}{N}}^{\frac{1}{N}}\times F_{R_{v}|r_{\kappa}}(r)\big|^{D_{r_\kappa,\theta_\kappa}^{\rm right}}_{D_{r_\kappa,\theta_\kappa}^{\rm left}}.
\end{align}  
For $m\in\{2,3,...,M\}$, $p_m^{v}$ depends only on the angle and $p_m^{\rm in}=p_m^{\rm out}$, which is given by
\begin{align}\label{eq:pm}
p_m^{v}& = F_{\Phi|\vartheta_\kappa}(\phi)\big|_{-\frac{m}{N}}^{-\frac{m-1}{N}} + F_{\Phi|\vartheta_\kappa}(\phi)\big|_{\frac{m-1}{N}}^{\frac{m}{N}}.
\end{align} 
By the law of total probability, $p_{M+1}^{v}= 1-\sum_{i=0}^{M}  p_i^{v}$.

\begin{IEEEproof}
\textcolor{black}{As discussed in Sec.~\ref{subsec:OrderedNode}, for an inner interferer located at $[\theta, r]^\mathsf{T}\in\Psi^{\kappa}_{\rm in}$, its distance satisfies $r<r_\kappa$.
Then, based on the definition of $p_0^{\rm in}$ and Proposition~\ref{Approx:NF-MLAP}, we obtain}
\begin{align}\label{eq:p0in}
    p_0^{\rm in}&=\mathbb{P}\left\{ \widetilde{\mathcal{G}}_{M}(\theta,r;\theta_\kappa,r_\kappa)=g_0 \big| r< r_\kappa \right\}
    \nonumber \\& =\textcolor{black}{\mathbb{P} \left\{ (\theta,r) \in \mathcal{R}_0 (\theta_\kappa,r_\kappa) \big| r < r_\kappa\right\}}
   \nonumber \\&=\mathbb{P} \left\{  |\phi| \le \frac{1}{N} \right\} \mathbb{P}\left\{  D_{r_\kappa,\theta_\kappa}^{\rm right}  \le r \le \infty \big| r < r_\kappa\right\}.
\end{align}
Using the conditional distribution of $\phi$ in \eqref{cdf:pdf:phi} and the conditional distribution of $r$ for $r<r_\kappa$ in \eqref{eq:Rin}, we can express \eqref{eq:p0in} explicitly as \eqref{eq:p0}.
Similarly, $p_0^{\rm out}$ can be obtained from the conditional distribution of $r$ for $r \ge r_\kappa$ in \eqref{eq:Rout}.
The derivations of the remaining probabilities $p_i^v$ follow analogously and are omitted here for brevity.
\end{IEEEproof}

\end{lemma}

Based on Theorem~\ref{them:cov} and Lemma~\ref{lemma:prob_Gain}, in the following, we provide the performance analysis under the NF-MLAP.
\begin{theorem} \label{theo:approxCP}
Under the NF-MLAP, given the $\kappa$-th active user at $[\theta_\kappa, r_\kappa]^{\mathsf{T}}$, the conditional \ac{CP} in \eqref{eq:condCP} can be expressed as
\begin{align}\label{eq:condCP_approx}
		&\widetilde{\rm CP}_{\kappa}(\tau |\theta_{\kappa},r_\kappa )
        \nonumber\\&= \textcolor{black}{\mathbb{P}\bigg\{ \sum_{\mathbf{u}\in \Psi_{\rm a} \setminus \{\mathbf{u}_\kappa\}}  \!\!\!\!\!\! \widetilde{\mathcal{G}}_{M} (\theta,r;\theta_\kappa,r_{\kappa}) < \frac{ 1 }{\tau} \bigg|~\theta_\kappa,r_\kappa \bigg\}}
        \\&
        =\frac{1}{2}- \frac{1}{2 {\rm j} \,\pi} \int_{0}^{\infty} \frac{1}{t} \, \big[ e^{-{\rm j}\, t \frac{1}{\tau}} \widetilde{\mathcal{L}}_{\theta_{\kappa},r_\kappa}(-{\rm j}\, t) -
        e^{{\rm j}\, t \frac{1}{\tau}} \widetilde{\mathcal{L}}_{\theta_{\kappa},r_\kappa}({\rm j}\, t)     \big] \mathrm{d}t.  \nonumber
\end{align}
Here, $\widetilde{\mathcal{L}}_{\theta_{\kappa},r_\kappa}(s)$ is the Laplace transform of the normalized aggregate interference under the NF-MLAP, which is given by
\begin{align}
    \widetilde{\mathcal{L}}_{\theta_{\kappa},r_\kappa}(s)=\big(\widetilde{\mathcal{L}}^{\rm in}_{\theta_{\kappa},r_\kappa}(s)\big)^{\kappa-1}\big(\widetilde{\mathcal{L}}^{\rm out}_{\theta_{\kappa},r_\kappa}(s)\big)^{N_{\rm a}-\kappa},
\end{align}
where 
\begin{align}
\widetilde{\mathcal{L}}^{v}_{\theta_{\kappa},r_\kappa}(s)
        = \sum_{i=0}^{M+1}p_{i}^{v} \exp(-s g_i), \text{ for } v\in\{\rm in,out\},
\end{align}
where $g_i$ is given in Proposition~\ref{Approx:NF-MLAP}, and $p_i^v$ is given in Lemma~\ref{lemma:prob_Gain}. 
Then, the overall \ac{CP} in \eqref{eq:overCP} under the NF-MLAP is
 \begin{align}\label{eq:CP_approx}
	\begin{split}
&\widetilde{\rm CP}_{\kappa} (\tau)
        \!=\! \int_{-\frac{\pi}{N_{\rm s}}}^{\frac{\pi}{N_{\rm s}}} \! \int_{0}^{R_{\rm c}} \!\!\! \widetilde{\rm CP}_{\kappa}(\tau |\theta_{\kappa},r_\kappa )  f_{R_{\kappa}}(r_\kappa) \frac{N_{\rm s}}{2\pi} \mathrm{d} r_\kappa \mathrm{d} \theta_{\kappa},
	\end{split}
\end{align}
where $f_{R_{\kappa}}(r_\kappa)$ is given in \eqref{eq:PDF_Rk}. 
	\begin{IEEEproof}
		 By replacing the actual \ac{NF} antenna pattern in \eqref{eq:antenna_pattern} with the proposed NF-MLAP in \eqref{eq:NF_MLAP} and following the derivation in Appendix~\ref{app:them:cov}, we complete the proof.
	\end{IEEEproof}
\end{theorem}

The \acp{CP} under the NF-MLAP, i.e., $\widetilde{\rm CP}_{\kappa}(\tau |\theta_{\kappa},r_\kappa)$ and $\widetilde{\rm CP}_{\kappa}(\tau) $, serves as approximations of the actual \acp{CP} under the exact pattern, i.e., ${\rm CP}_{\kappa}(\tau |\theta_{\kappa},r_\kappa)$ and ${\rm CP}_{\kappa}(\tau) $. Furthermore, the approximations of the conditional SE, overall SE, and ASE can be obtained by replacing ${\rm CP}_{\kappa}(\tau |\theta_{\kappa},r_\kappa)$ and ${\rm CP}_{\kappa}(\tau) $ with $\widetilde{\rm CP}_{\kappa}(\tau |\theta_{\kappa},r_\kappa)$ and $\widetilde{\rm CP}_{\kappa}(\tau) $ in \eqref{eq:condAR}-\eqref{eq:ASE}, denoted as $\widetilde {\rm SE}_{\kappa}(\tau |\theta_{\kappa},r_\kappa)$, $\widetilde{\rm SE}_{\kappa} (\tau)$, and $\widetilde{\rm ASE}_{\kappa} (\tau)$, respectively. 

From Lemma~\ref{lemma:prob_Gain}, given $[\theta_\kappa, r_\kappa]^{\mathsf{T}}$ and $N$, the probabilities $\{p_i^v\}$ can be computed offline. 
Therefore, the NF-MLAP enables a closed-form approximation for the Laplace transform of the interference $I_\kappa$ in Theorem~\ref{theo:approxCP}.
This indicates that the numerical evaluation of the \acp{CP}, \acp{SE}, and \ac{ASE} under the proposed NF-MLAP is significantly simpler than that under the exact antenna pattern. 

\vspace{-2mm}

\subsection{Performance Upper Bounds Under NF-MLAP}\label{sec:upper}

We see from Theorem~\ref{theo:approxCP} that even when the Laplace transform of the interference is available in closed form, computing the conditional \ac{CP} in \eqref{eq:condCP_approx} still requires a single integration. 
To further reduce the computational complexity, we next derive a closed-form upper bound of $\widetilde{\rm CP}_{\kappa} (\tau|\theta_\kappa,r_\kappa)$, which is denoted by $\widehat{\rm CP}_{\kappa}(\tau |\theta_{\kappa},r_\kappa )$. 
\textcolor{black}{Note that $\widetilde{\rm CP}_{\kappa}(\tau |\theta_\kappa,r_\kappa)$ in \eqref{eq:condCP_approx} requires that the normalized aggregate interference gain from all interferers is below $\frac{1}{\tau}$.
Instead of directly handling this sum, we upper bound it using a sufficient condition that each interferer contributes an interference gain that is below $\frac{1}{\tau}$.}

\begin{theorem}\label{theo:approxCP_upper}
Under the NF-MLAP, $\widetilde{\rm CP}_{\kappa}(\tau |\theta_{\kappa},r_\kappa )$ is upper bounded by \textcolor{black}{the probability that all interferers individually satisfy the interference constraint $g_i < \frac{1}{\tau}$,} i.e., 
\begin{align}\label{eq:approx_CP_cond_upper}
    &\widetilde{\rm CP}_{\kappa}(\tau |\theta_{\kappa},r_\kappa )  \le \quad\widehat{\rm CP}_{\kappa}(\tau |\theta_{\kappa},r_\kappa )
    \\&=\textcolor{black}{\prod_{\mathbf{u} \in \Psi_{\rm a} \setminus \{\mathbf{u}_\kappa\}}  
 \mathbb{P}\left\{ \widetilde{\mathcal{G}}_{M}(\theta ,r ;\theta_\kappa,r_{\kappa}) < \frac{ 1 }{\tau} ~\bigg|~\theta_\kappa,r_\kappa\right\}  }
    \nonumber\\&=\left(\sum_{i=0}^{M+1}\!\mathbbm{1}\!\left(g_i <\frac{1}{\tau}\right) p_i^{\rm in }\right)^{\!\!\!\kappa-1}
     \!\!\!\left(\sum_{i=0}^{M+1}\!\mathbbm{1}\!\left(g_i<\frac{1}{\tau}\right) p_i^{\rm out }\right)^{\!\!N_{\rm a}-\kappa}  \!\!\!\!\!\!\!\!\!\!\!\!,   \nonumber
\end{align}
where $g_i$ and $p_i$ are given in Proposition~\ref{Approx:NF-MLAP} and Lemma~\ref{lemma:prob_Gain}, respectively, and $\mathbbm{1}(\cdot)$ is the indicator function.

\begin{IEEEproof}
		See Appendix~\ref{app:theo:approxCP_upper}.
	\end{IEEEproof}
\end{theorem}

By replacing $\widetilde{\rm CP}_{\kappa}(\tau |\theta_{\kappa},r_\kappa )$ with $\widehat{\rm CP}_{\kappa}(\tau |\theta_{\kappa},r_\kappa  )$, we obtain the upper bounds of $\widetilde{\rm CP}_{\kappa}(\tau )$, $\widetilde {\rm SE}_{\kappa}(\tau |\theta_{\kappa},r_\kappa)$, $\widetilde{\rm SE}_{\kappa} (\tau)$, and $\widetilde{\rm ASE}_{\kappa} (\tau)$ in \eqref{eq:condAR}-\eqref{eq:ASE}, \textcolor{black}{denoted by
$\widehat{\rm CP}_{\kappa}(\tau )$, $\widehat {\rm SE}_{\kappa}(\tau |\theta_{\kappa},r_\kappa)$, $\widehat{\rm SE}_{\kappa} (\tau)$, and $\widehat{\rm ASE}_{\kappa} (\tau)$, respectively.}
Since $\widehat{\rm CP}(\tau |\theta_{\kappa},r_\kappa )$ is in closed form, evaluating these upper bounds becomes computationally efficient.
\color{black}
We next characterize a sufficient condition under which the upper bound becomes exact. 

\begin{cor}\label{cor:upper_eq}
A sufficient condition that the equality in \eqref{eq:approx_CP_cond_upper} holds is ${\tau}\ge\tau^*(M)$, where $\frac{1}{\tau^*(M)}= \min_{i\in\{0,1,...,M\}}{g_i}$,
i.e., the minimum non-zero gain level in the NF-MLAP is greater than $\frac{ 1 }{\tau}$. In this case, 
\begin{align}\label{eq:approx_CP_const}
\widetilde{\rm CP}_{\kappa}(\tau |\theta_{\kappa},r_\kappa )  &= 
     \widehat{\rm CP}_{\kappa}(\tau |\theta_{\kappa},r_\kappa  )
    \\&=\left( p_{M+1}^{\rm in } \right)^{\kappa-1} \left(  p_{M+1}^{\rm out } \right)^{N_{\rm a}-\kappa}, \text{ for } \tau \ge \tau^*(M).\nonumber
\end{align}
    \begin{IEEEproof}
		See Appendix~\ref{app:cor:upper_eq}.
	\end{IEEEproof}
\end{cor}

The condition in Corollary~\ref{cor:upper_eq} corresponds to a stringent SIR requirement under which even the weakest non-zero gain defined in the NF-MLAP satisfies $g_i\ge \frac{1}{\tau}$. As a result, any interferer located in the regions associated with non-zero gain levels, i.e., $\{\mathcal{R}_i(\theta_\kappa,r_\kappa)\}_{i=0}^{M}$, is sufficient to violate the SIR constraint, \textcolor{black}{making the per-interferer condition in \eqref{eq:approx_CP_cond_upper} equivalent to the aggregate interference condition in \eqref{eq:condCP_approx} and rendering the upper bound tight.}

 \vspace{-2mm}
 
\subsection{Choice of $M$ for Performance Evaluation}\label{subsec:M}

From Lemma~\ref{lemma:prob_Gain}, once $[\theta_\kappa,r_\kappa]^{\mathsf{T}}$, $M$, and $N$ are given, the value of $p_{M+1}^{v}$ becomes constant.  
Corollary~\ref{cor:upper_eq} implies that when $\min_{i\in\{0,1,...,M\}}{g_i}=g_M$, the condition $\tau\ge\frac{1}{g_M}$ leads to a constant value for both $\widehat{\rm CP}_{\kappa}(\tau |\theta_{\kappa},r_\kappa )$ and $\widetilde{\rm CP}_{\kappa}(\tau |\theta_{\kappa},r_\kappa )$, regardless of further increases in $\tau$.  
However, in practice, the \ac{CP} decreases with $\tau$ and eventually approaches zero. This mismatch implies that, for the given level $M$, the NF-MLAP may fail to accurately characterize interference when $\tau > \tau^*(M)$. 

To better understand this behavior, we characterize the coverage performance in terms of dominant interferers. 
From \eqref{eq:approx_CP_cond_upper}, the CP is dominated by interferers located in spatial regions where the associated non-zero gain exceeds $\frac{1}{\tau}$, which violates the SIR constraint.  
Under the NF-MLAP, the antenna pattern is approximated by retaining the beamfocusing region together with the first $(M-1)$ sidelobe regions, while all remaining regions are grouped into a zero-gain region.  
However, if sidelobe regions with gains exceeding $\frac{1}{\tau}$ are not included in the NF-MLAP, then the corresponding dominant interferers are not captured by the approximation, leading to a mismatch between the approximate and actual system behaviors. 

This observation suggests that, to ensure accurate interference characterization, $M$ should be chosen such that all sidelobe gain levels satisfying $g_i \ge \frac{1}{\tau}$ are included in the NF-MLAP. A sufficient condition for this is $g_M < \frac{1}{\tau}$. 
We define the corresponding minimum value of $M$ as $M^*(\tau)$. 
\color{black}  
We next characterize how $M^*(\tau)$ scales with $\tau$ when sidelobe
regions need to be retained, i.e., for $M^*(\tau)\ge2$.  

\begin{prop}\label{prop:M_scaling}
For a given $\tau$, a sufficient condition to ensure that the NF-MLAP accurately captures dominant interferers is $M \ge M^*(\tau)$. For $M^*(\tau)\ge2$,  
\begin{align}\label{eq:M-tau}
M^*(\tau) \approx \frac{1}{2}\left(\frac{\sqrt{2\delta \tau}}{\pi} + 1 \right). 
\end{align}
where $\delta=\frac{1}{\sqrt2}$ is defined in \eqref{eq:GAapprox}.
This implies that $M^*(\tau)=\mathcal{O}(\sqrt{\tau})$, where $\tau$ is expressed in linear scale. 
\begin{proof}
From \eqref{eq:GAapprox}, the $m$-th gain level is given by 
\begin{align}\label{eq:gm_exact}
g_m = \frac{\delta}{2} \frac{\sin^2\!\left(\pi N \frac{2m-1}{2N}\right)}{N^2 \sin^2\!\left(\pi \frac{2m-1}{2N}\right)},~ m\ge 2.
\end{align}
For large $N$ and moderate $m$, by applying the approximation $\sin x \approx x$ to \eqref{eq:gm_exact}, $g_m$ can be approximated as
\begin{align}\label{eq:gm_approx}
    g_m \approx \frac{2\delta}{\pi^2 (2m-1)^2}, ~m\ge 2.
\end{align}
\color{black} 
Substituting \eqref{eq:gm_approx} into $g_M < \frac{1}{\tau}$, we can obtain \eqref{eq:M-tau}.  
\end{proof}
\end{prop}
\color{black}

The above result reflects that, as $\tau$ increases, interferers in more sidelobe regions can lead to violations of the SIR constraint. As a result, these regions need to be retained in the NF-MLAP, leading to the scaling $M=\mathcal{O}(\sqrt{\tau})$. 

 \color{black}
 
 \vspace{-2mm} 
 
\subsection{Extension to SINR-related Performance Analysis}\label{subsec:SINR}
Based on Sec.~\ref{sec:signal}, the \ac{SINR} of the $\kappa$-th active user is
 \begin{align}\label{eq:SINR_k}
    {\rm SINR}_\kappa&=\frac{P_{\kappa,{\rm S}}}{P_{\kappa,{\rm I}}+\sigma^2} 
   = \frac{1}{ I_\kappa+ \frac{N_{\rm a} \sigma^2}{ P_{\rm t} N\zeta r_\kappa^{-\alpha}}   } .
\end{align}
The SINR-related conditional CP, ${\rm CP}^{\rm sinr}_{\kappa}(\tau |\theta_{\kappa},r_\kappa )=\mathbb{P}\{{\rm SINR}_{\kappa}>\tau ~|~\theta_{\kappa},r_\kappa \}$, is given below.
\begin{cor}\label{cor:SINR}
    Given the $\kappa$-th active user at $[\theta_{\kappa},r_\kappa]^{\mathsf{T}}$, the SINR-related conditional CP is
\begin{align}\label{eq:SINR-condCP}
&{\rm CP}^{\rm sinr}_{\kappa}(\tau |\theta_{\kappa},r_\kappa ) = {\rm CP}_{\kappa}\bigg(\frac{1}{\frac{1}{\tau} -  \frac{N_{\rm a} \sigma^2}{ P_{\rm t} N\zeta r_\kappa^{-\alpha}}} ~\bigg|~\theta_{\kappa},r_\kappa \bigg),
\end{align}
where ${\rm CP}_{\kappa}(\cdot|\cdot)$ is the SIR-related conditional CP given in \eqref{eq:CP_exact}. 
\begin{proof}
From the definition of the conditional CP, we have
           \begin{align}
&{\rm CP}^{\rm sinr}_{\kappa}(\tau |\theta_{\kappa},r_\kappa )=\mathbb{P}\left\{\frac{1}{I_\kappa+ \frac{N_{\rm a} \sigma^2}{ P_{\rm t} N\zeta r_\kappa^{-\alpha}}   } >\tau ~|~\theta_{\kappa},r_\kappa \right\}
\nonumber\\&=\mathbb{P}\left\{  I_\kappa  <  \frac{1}{\tau} -  \frac{N_{\rm a} \sigma^2}{ P_{\rm t} N\zeta r_\kappa^{-\alpha}} ~|~\theta_{\kappa},r_\kappa \right\}.
\end{align}
Following the same derivation steps as in Appendix~\ref{app:them:cov}, we obtain~\eqref{eq:SINR-condCP}, which completes the proof. 
\end{proof}
\end{cor} 
From Corollary~\ref{cor:SINR}, the SINR-related performance metrics can be directly obtained from their SIR counterparts by substituting the SIR threshold $\tau$ with $\left(\frac{1}{\tau} -  \frac{N_{\rm a} \sigma^2}{ P_{\rm t} N\zeta r_\kappa^{-\alpha}}\right)^{-1}$. Specifically, by applying this substitution in Theorem~\ref{theo:approxCP}, Theorem~\ref{theo:approxCP_upper} and~\eqref{eq:condAR}-\eqref{eq:ASE}, we obtain the approximations of the SINR-related performance metrics and their upper bounds.

\color{black}
 \vspace{-2mm}
 
\section{Analytical Insights and Scaling Laws}\label{sec:scaling_insight}

In this section, we extract analytical insights into the scaling laws of the \ac{ASE} with respect to the antenna number $N$ and the number of active users $N_{\rm a}$. 
\textcolor{black}{The analysis is conducted under the NF-MLAP with a given discretization level $M$} and builds on the performance upper bound in Theorem~\ref{theo:approxCP_upper}.
 
 \vspace{-2mm}
 
\subsection{Critical Interference Region-Based CP}

From the first equality in \eqref{eq:approx_CP_cond_upper}, the upper bound of the conditional \ac{CP} can be reformulated as
\begin{align} \label{eq:CP_danger}
    \widehat{\rm CP}_{\kappa}\!(\tau |\theta_{\kappa},r_\kappa ) 
    \!=\!\!\!\! \!\!\!\!\!\prod_{\mathbf{u} \in \Psi_{\rm a} \setminus \{\mathbf{u}_\kappa\}} \!\!\! \!\!\!\!\!\!
 \big(1\!-\!\underbrace{\mathbb{P}\left\{  (\theta,r) \in \mathcal{R}_{\tau} (\theta_\kappa,r_\kappa) \right\}}_{P_\kappa(\tau|\theta_\kappa,r_\kappa)} \big), 
\end{align}
where $\mathcal{R}_{\tau} (\theta_\kappa,r_\kappa)=\{(\theta,r): \widetilde{\mathcal{G}}_{M}(\theta ,r ;\theta_\kappa,r_{\kappa}) \ge \frac{ 1 }{\tau}\}$ defines the \emph{critical interference region}, 
i.e., the set of interferer locations where the interference gain exceeds $\frac{1}{\tau}$, such that a single interferer can individually violate the SIR constraint.
Under the NF-MLAP, this region can be explicitly characterized as
\begin{align}\label{eq:R_danger}
    \mathcal{R}_{\tau} (\theta_\kappa,r_\kappa)=\bigcup_{i,g_i\ge\frac{1}{\tau}} \mathcal{R}_{i} (\theta_\kappa,r_\kappa),
\end{align} 
where $\mathcal{R}_{i} (\theta_\kappa,r_\kappa)$ is the $i$-th discretized region associated with the quantized gain level $g_i$, as defined in Proposition~\ref{Approx:NF-MLAP}.
Accordingly, $P_\kappa(\tau|\theta_\kappa,r_\kappa)$ can be interpreted as the probability that a randomly located interferer falls into the critical interference region.
This reformulation shows that $\widehat{\rm CP}_{\kappa}(\tau |\theta_{\kappa},r_\kappa )$ is determined by the probability that all interferers lie outside the critical interference region.
Therefore, the scaling behavior of the conditional CP can be characterized by analyzing how the size of $\mathcal{R}_{\tau}(\theta_\kappa,r_\kappa)$ varies with $N$.
 
 \vspace{-2mm}
 
\subsection{Scaling of Conditional CP}\label{subsec:low-high-tau}

From \eqref{eq:R_danger}, the critical interference region consists of the NF-MLAP regions whose associated gain levels satisfy $g_i \ge \frac{1}{\tau}$ and therefore depends on the SIR threshold $\tau$.
As a result, different NF-MLAP regions dominate $P_\kappa(\tau|\theta_\kappa,r_\kappa)$ under different $\tau$, leading to the following two representative regimes.


\subsubsection{Low-Threshold Regime}
When $\tau$ is sufficiently small such that only the strongest gain level $g_1$ exceeds $\frac{1}{\tau}$, the critical interference region reduces to the beamfocusing region, i.e., $\mathcal{R}_{\tau} (\theta_\kappa,r_\kappa)
=
\mathcal{R}_{1} (\theta_\kappa,r_\kappa)$. 
This region is jointly characterized by the finite beam width and beam depth in the \ac{NF}.
From Sec.~\ref{subsec:BW}, the finite beam width scales as ${\rm BW} = \mathcal{O}(N^{-1})$.
From \eqref{eq:BD} in Sec.~\ref{subsec:BD}, the finite beam depth scales as
${\rm BD}(\theta_\kappa,r_\kappa)
= \mathcal{O}(N^{-2}).$
Hence, the size of the beamfocusing region scales as
$|\mathcal{R}_{\tau}(\theta_\kappa,r_\kappa)|
= \mathcal{O}(N^{-1}) \cdot \mathcal{O}(N^{-2})
= \mathcal{O}(N^{-3})$.
This implies
\begin{align}\label{eq:O(N3)}
P_\kappa(\tau|\theta_\kappa,r_\kappa)
= \mathcal{O}(N^{-3}), \text{ for low } \tau.
\end{align}
Substituting \eqref{eq:O(N3)} into \eqref{eq:CP_danger}, we obtain 
\begin{align} \label{eq:CP_danger_low}
   &\widehat{\rm CP}_{\kappa}(\tau|\theta_{\kappa},r_\kappa ) \nonumber\\&=  \prod_{\mathbf{u} \in \Psi_{\rm a} \setminus \{\mathbf{u}_\kappa\}} \!\!\! \!\!\!\!\! \big(1-\mathcal{O}(N^{-3})\big)=\big(1-\mathcal{O}(N^{-3})\big)^{N_{\rm a}-1} \\&\overset{(a)}{\approx} 
1-\mathcal{O}\!\left((N_{\rm a}-1)N^{-3}\right)=
1-\mathcal{O}\!\left(N_{\rm a} N^{-3}\right), \text{for low } \tau,\nonumber
\end{align}
where (a) uses the first-order approximation $(1-x)^n \approx 1-nx$ for $x\ll 1$.  
This indicates that, under a relaxed SIR requirement, the coverage performance is mainly determined by interferers located within the NF beamfocusing region, whose size shrinks rapidly with $N$.

\subsubsection{High-Threshold Regime}
\textcolor{black}{When $\tau$ increases, the critical interference region expands to include some sidelobe regions as $\mathcal{R}_{\tau} (\theta_\kappa,r_\kappa)
=
\bigcup_{i=0}^{m} \mathcal{R}_{i} (\theta_\kappa,r_\kappa)$, for some $m\in[2,M]$, i.e., at least one sidelobe gain level satisfies $g_m \ge \frac{1}{\tau}$.}
Since each sidelobe region occupies an angular interval of width $\mathcal{O}(N^{-1})$, 
$|\mathcal{R}_{\tau}(\theta_\kappa,r_\kappa)|=\mathcal{O}(N^{-1})$, which yields 
\begin{align}
P_\kappa(\tau|\theta_\kappa,r_\kappa)
=
\mathcal{O}(N^{-1}), \text{ for high } \tau.
\end{align}
Following the same steps in \eqref{eq:CP_danger_low}, we obtain
\begin{align}
\label{eq:CP_danger_high}
\widehat{\rm CP}_{\kappa}(\tau|\theta_{\kappa},r_\kappa ) 
\approx
1-\mathcal{O}\!\left(N_{\rm a}N^{-1}\right), \text{ for high } \tau.
\end{align}
\textcolor{black}{This shows that, under a stringent SIR requirement, the coverage performance is affected by a much broader set of interferers, including those located in sidelobe regions whose size decreases slowly with $N$.}

\color{black}
To provide a more concrete interpretation of the low- and high-threshold 
regimes, we further evaluate the condition $g_i \ge \frac{1}{\tau}$ based on the gain levels defined in the NF-MLAP.
Specifically, from \eqref{eq:GAapprox}, $\tau\ge\frac{1}{g_1}\approx 4.5~\rm dB$; using the approximation in \eqref{eq:gm_approx}, $\tau\ge\frac{1}{g_2}\approx 18~\rm dB$ and $\tau\ge\frac{1}{g_3}\approx 22.4~\rm dB$.
This indicates that the transition to the sidelobe-involved 
high-threshold regime occurs around $\tau \approx 18~\rm dB$, at which the first sidelobe region becomes part of the critical interference region.

\color{black}

 \vspace{-2mm}
 
\subsection{ASE Scaling and Design Insights}\label{subsec:ASEscale}

According to \eqref{eq:overCP}, averaging the conditional CP over the spatial distribution of the tagged user gives the overall \ac{CP}. Since this averaging operation does not change the scaling order in \eqref{eq:CP_danger_low} and \eqref{eq:CP_danger_high}, the overall CP behaves as 
\begin{align}\label{eq:CP_danger_low_high}
\widehat{\rm CP}_{\kappa}(\tau)
&\approx\begin{cases}
 1- c_{\tau_{\rm low}}N_{\rm a}N^{-3}, &  \text{for low } \tau, 
 \\
 1- c_{\tau_{\rm high}} N_{\rm a}N^{-1} ,
&  \text{for high } \tau,
\end{cases} 
\end{align}
where $c_{\tau_{\rm low}}$ and $c_{\tau_{\rm high}}$ are constants
independent of $N$ and $N_{\rm a}$.
Then, based on the definition of the ASE in \eqref{eq:ASE}, we obtain 
\begin{align}\label{eq:ase_scale}
\widehat{\rm ASE} (\tau) 
&=
\frac{N_{\rm s}}{\pi R_{\rm c}^2}
\sum_{\kappa=1}^{N_{\rm a}}
\widehat{\rm CP}_{\kappa}(\tau)\log_2(1+\tau)
\nonumber\\&\approx\begin{cases}
 C_\tau N_{\rm a}
\left(
1- c_{\tau_{\rm low}}N_{\rm a}N^{-3}
\right), &  \text{for low } \tau, 
 \\
 C_\tau N_{\rm a}
\left(
1- c_{\tau_{\rm high}}N_{\rm a}N^{-1}
\right),
&  \text{for high } \tau,
\end{cases}  
\end{align}
where $C_\tau=\frac{N_{\rm s}}{\pi R_{\rm c}^2}\log_2(1+\tau)$ is independent of $N$ and $N_{\rm a}$. 
We can further decompose the ASE in \eqref{eq:ase_scale} as  
\begin{align}\label{eq:ase_mu}
\widehat{\mathrm{ASE}}(\tau) 
\approx
\underbrace{C_\tau N_{\rm a}}_{ \text{spatial reuse gain}}
-
\underbrace{ C_\tau  c_{\tau_{\mu}} N_{\rm a}^2 N^{\eta_\mu}}_{ \text{interference penalty}}
~~,
\end{align}
where $\mu\in\{\rm low,high\}$, $\eta_{\rm low}=-3$ for low $\tau$, and $\eta_{\rm high}=-1$ for high $\tau$.  
Eq.~\eqref{eq:ase_mu} reveals that the ASE scaling is fundamentally determined by the interplay between spatial reuse gain and inter-user interference penalty. 
The resulting scaling laws under low and high SIR thresholds provide useful design insights, as discussed below. 

\subsubsection{Scaling With Antenna Number $N$}\label{subsubsec:scale-N}
From \eqref{eq:ase_mu}, the ASE approaches the spatial reuse limit $C_\tau N_{\rm a}$ as $N$ increases, which corresponds to the regime where the inter-user interference becomes negligible.
The gap to this limit is governed by the interference penalty, which decays as $\mathcal{O}(N^{-3})$ for low $\tau$ and $\mathcal{O}(N^{-1})$ for high $\tau$.  
Therefore, the benefit of increasing $N$ is fundamentally tied to how effectively NF beamfocusing reduces the size of the critical interference region, thereby suppressing the interference penalty.
For low SIR thresholds, since the interference penalty decays rapidly with $N$, the system quickly approaches the interference-negligible regime, 
and achieves most of the spatial reuse gain with a moderate antenna size. 
Beyond this point, further increasing $N$ yields only marginal performance improvement, i.e., the ASE tends to saturate once the dominant interference is effectively suppressed. 
In contrast, for high SIR thresholds, the system remains interference-limited over a wider range of antenna sizes due to the slower decay of the interference penalty with $N$, and thus requires a substantially larger antenna array to approach the spatial reuse limit.

\subsubsection{Scaling With Active-User Number $N_{\rm a}$}
\label{subsubsec:scale-Na}
For the number of active users, from \eqref{eq:ase_mu}, there exists a fundamental tradeoff between the {\em linear} growth of the spatial reuse gain and the {\em quadratic} growth of the interference penalty. 
Hence, the ASE generally exhibits a unimodal behavior with respect to $N_{\rm a}$.
In the low-threshold regime, the interference penalty scales as $\mathcal{O}(N_{\rm a}^2 N^{-3})$ and is strongly suppressed for large $N$. This suggests that the penalty grows much more slowly than the spatial reuse gain.  
As a result, the ASE increases steadily with $N_{\rm a}$ over a broad practical range of $N_{\rm a}\le N$.
In contrast, in the high-threshold regime, the interference penalty scales as $\mathcal{O}(N_{\rm a}^2 N^{-1})$ and becomes much more pronounced. As a result, the ASE increases initially with $N_{\rm a}$ due to the spatial reuse gain, but then decreases rapidly once the interference penalty dominates. 
Balancing the two terms yields an optimal number of active users that scales linearly with $N$, implying an approximately constant optimal user-to-antenna ratio $\frac{N_{\rm a}}{N}$. 
This further indicates that the admissible number of active users is fundamentally constrained by the SIR requirement: Relaxed thresholds allow larger user loading with mild interference penalty, whereas stringent thresholds require careful control of $N_{\rm a}$ to avoid severe performance degradation.

The above analysis reveals that the additional distance-domain selectivity of NF beamfocusing is particularly beneficial under relaxed SIR requirements, where the suppression of strong interference within the localized beamfocusing region enables efficient spatial reuse. 
Under stringent SIR requirements, however, this benefit becomes less pronounced, since the dominant interference is no longer confined to the beamfocusing region but extends over a much larger spatial area. 
As a result, the system becomes more sensitive to inter-user interference, which limits the number of simultaneously served users for achieving high \ac{ASE}.
 

\color{black}

 \vspace{-2mm}

\section{Results and Discussions}\label{sec:results}


 \begin{table}[t!]\caption{Default values of system parameters.}
\centering
\begin{center}
{\linespread{0.95}
\renewcommand{\arraystretch}{1.2} 
    \begin{tabular}{ | {c} | {c} || {c} | {c} | }
    \hline
        \hline
    {Parameters} & {Values} & {Parameters} & {Values} \\ \hline 
    $N$ & $256$ &  $N_{\rm a}$ & $15$ \\ \hline
    $f_{\rm c}$ & $28~\rm GHz$ & $R_{\rm c}$ & $150~\rm m$ \\ \hline 
     {$\beta_\gamma$} &  $1.3$   & $M$ & $10$ \\ \hline
    $P_{\rm t}$ & $10~\rm W$ & $\alpha$ & $2$ \\ \hline
    \end{tabular}} 
\end{center}
\label{tab:simulation}
\end{table}

\begin{figure*}[t!]
\begin{minipage}{.325\textwidth}
    \centering
    \includegraphics[width=1.1\linewidth]{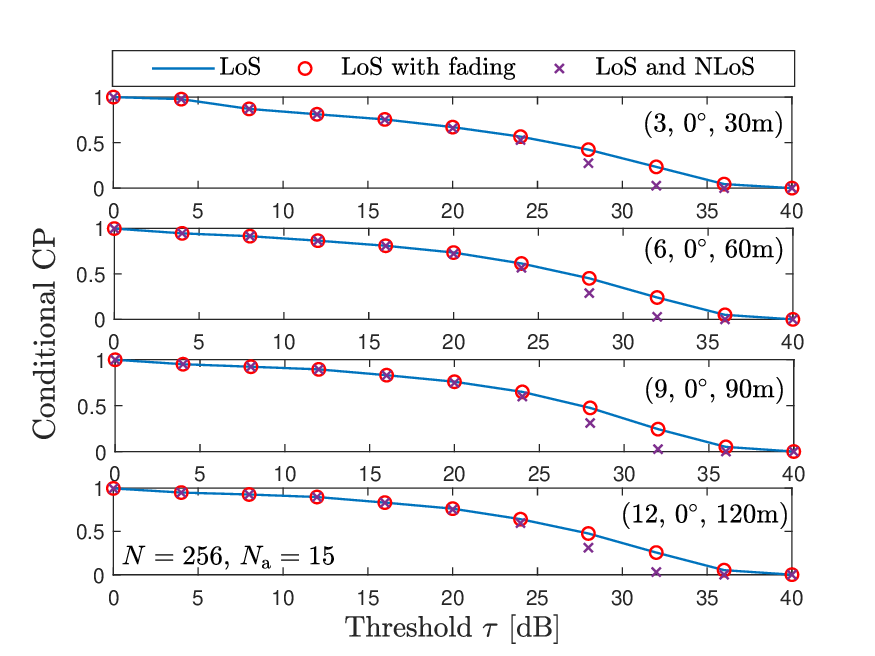}  
    \caption{\textcolor{black}{Impact of small-scale fading and NLoS paths on SIR-related conditional CPs for the $\kappa$-th user at $[\theta_\kappa,r_\kappa]^{\mathsf{T}}$, denoted as $(\kappa,\theta_\kappa,r_\kappa)$.}}
    \label{fig:condCPsir_LoS}
\end{minipage}
\hfill
\begin{minipage}{.325\textwidth}
    \centering
    \includegraphics[width=1.1\linewidth]{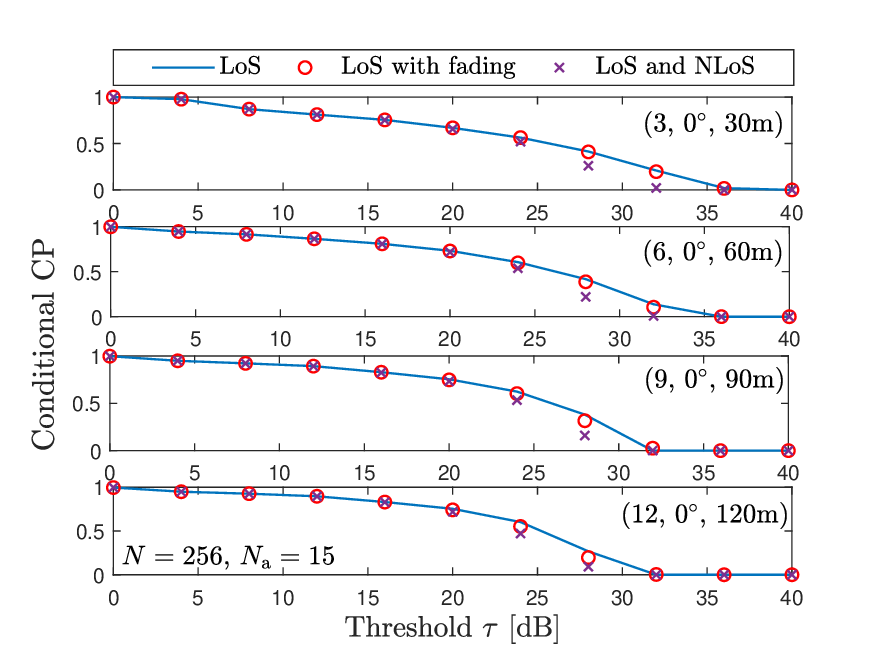}
    \caption{\textcolor{black}{Impact of small-scale fading and NLoS paths on SINR-related conditional CPs for the $\kappa$-th user at $[\theta_\kappa,r_\kappa]^{\mathsf{T}}$, denoted as $(\kappa,\theta_\kappa,r_\kappa)$.}}
    \label{fig:condCPsinr_LoS}
    \end{minipage}
\hfill
\begin{minipage}{.325\textwidth}
	\centering
    \includegraphics[width=1.1\linewidth]{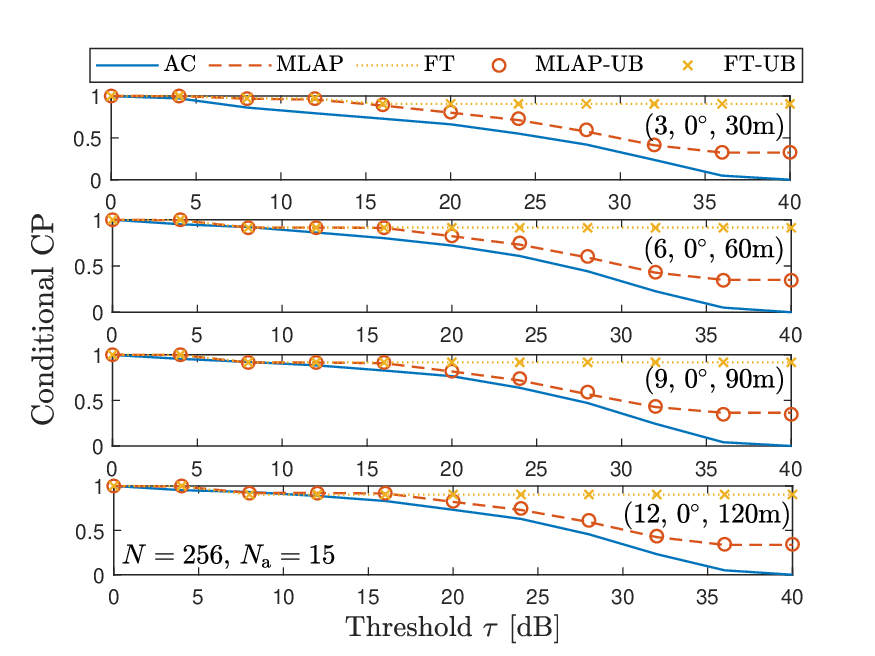} 
    \caption{SIR-related conditional CP under different antenna patterns for the $\kappa$-th user at $[\theta_\kappa,r_\kappa]^{\mathsf{T}}$, denoted as $(\kappa,\theta_\kappa,r_\kappa)$.}
    \label{fig:condCP} 
    \end{minipage}
\end{figure*}
 
This section presents numerical evaluations of the performance of the \ac{ELAA}-assisted NF multi-user communication. 
The default values of system parameters are summarized in Table~\ref{tab:simulation}, unless otherwise specified. The abbreviations of different antenna patterns are defined as follows: 
\begin{itemize}
    \item \textbf{AC}: Monte-Carlo simulation results obtained using the actual antenna pattern in \eqref{eq:antenna_pattern}. 
    \item \textbf{MLAP}: Numerical results based on Theorem~\ref{theo:approxCP} using the proposed NF-MLAP with $M=10$.  
    \item \textbf{FT}: Numerical results based on Theorem~\ref{theo:approxCP} using the proposed NF-MLAP with $M=1$, where the angular-domain pattern is approximated as a flat-top pattern.
    \item \textbf{MLAP-UB}: Numerical results based on Theorem~\ref{theo:approxCP_upper}, representing the upper bound of the `MLAP' curve. 
    \item \textbf{FT-UB}: Numerical results based on Theorem~\ref{theo:approxCP_upper}, representing the upper bound of the `FT' curve. 
\end{itemize}

\vspace{-2mm}

\subsection{Impact of Small-Scaling Fading and NLoS Component}\label{subsec:fading-NLoS}

\color{black}


We examine the validity of the channel model adopted in Sec.~\ref{subsec:channel}.
Specifically, the `LoS' model corresponds to the adopted channel model.
To account for the small-scale fading on the LoS path, we consider a `LoS with fading' model, where the LoS component is multiplied by a Nakagami-m random coefficient with the shaping parameter being $3$~\cite{ActualAntenna,FlatTop}.
To further account for multipath propagation, we consider a `LoS and NLoS' model, where the `LoS with fading' channel is superimposed with a NLoS component modeled as a circularly symmetric complex Gaussian random variable~\cite{yuan2024scalable}. Since the channel gains of NLoS paths are typically $20~\rm dB$ weaker than those of LoS paths in high-frequency systems~\cite{20dB}, we set the Rician K-factor to $20~\rm dB$.

Using the actual antenna pattern, Figs.~\ref{fig:condCPsir_LoS} and \ref{fig:condCPsinr_LoS} compare the simulation results of the SIR- and SINR-related conditional CPs under these three channel models, respectively. We observe that the curves under the `LoS' and `LoS with fading' models almost overlap over the entire threshold range. This indicates that the impact of small-scale fading on the LoS component is negligible.
Moreover, the `LoS' and `LoS and NLoS' curves remain very close across most thresholds, and only a slight deviation is observed in the high-threshold regime (e.g., at $28~\rm dB$ and $32~\rm dB$). 
This indicates that the contribution of the NLoS component is negligible in most operating regimes due to its much weaker power compared with the LoS component.
These results support the use of the simplified LoS channel model in our analysis.

\color{black}
 \vspace{-2mm}
 
\subsection{Accuracy of NF-MLAP Based Performance Approximation}\label{subsec:simu-NF-MLAP}

\begin{figure*}[t!]
\begin{minipage}{.325\textwidth}
  \vspace{-3mm}
    \centering
    \includegraphics[width=1.1\linewidth]{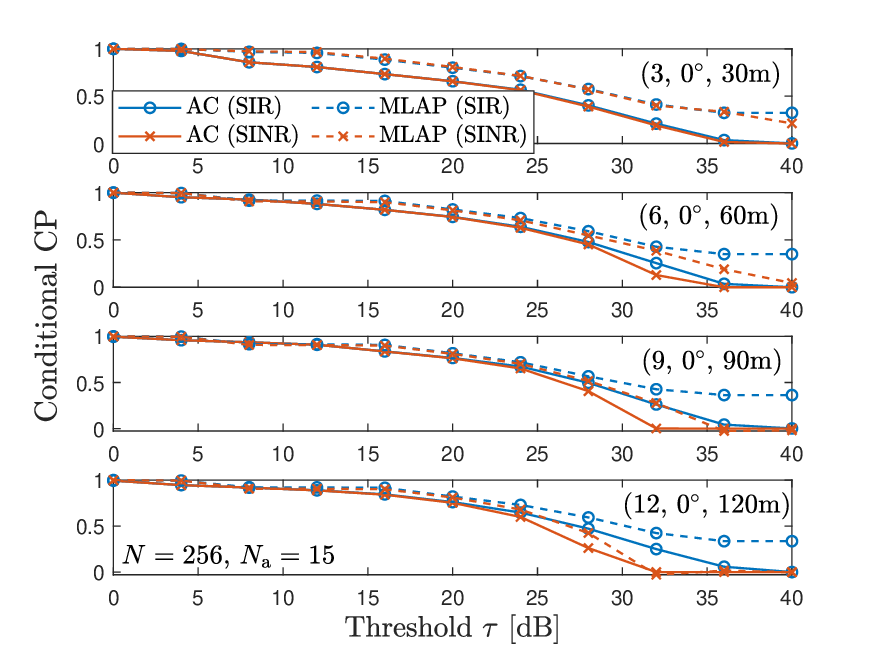} 
    \vspace{-6mm}
    \caption{Comparison of SIR- and SINR-related conditional CPs for the $\kappa$-th user at $[\theta_\kappa,r_\kappa]^{\mathsf{T}}$, denoted as $(\kappa,\theta_\kappa,r_\kappa)$.}
    \label{fig:condCPsinr}
\end{minipage}
\hfill
\begin{minipage}{.325\textwidth}
  \vspace{-3mm}
    \centering
    \includegraphics[width=1.1\linewidth]{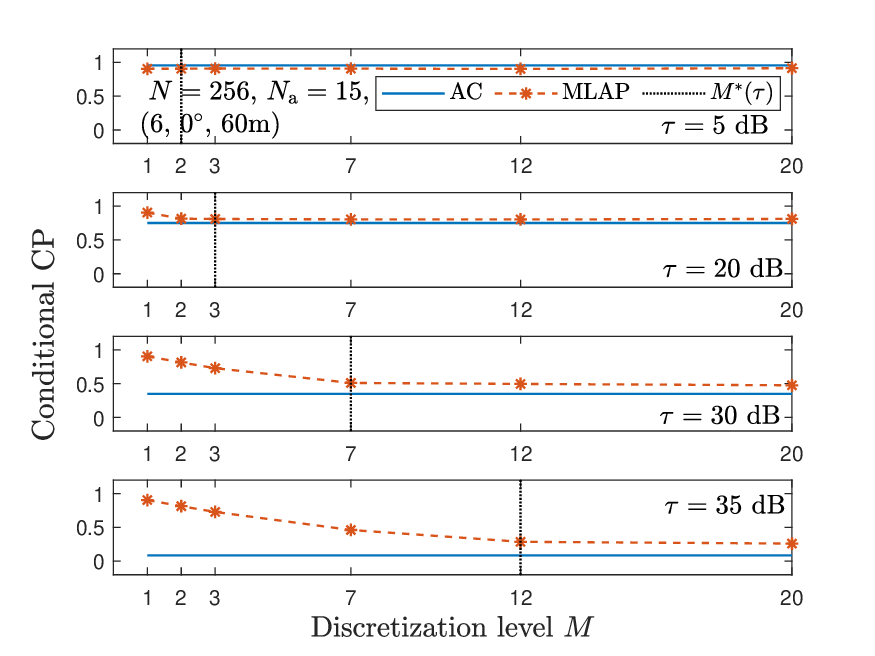}
    \vspace{-6mm}
    \caption{\textcolor{black}{Approximation accuracy of the SIR-related conditional CP under the NF-MLAP with different discretization levels $M$.}}
    \label{fig:value-M}
    \end{minipage}
\hfill
\begin{minipage}{.325\textwidth}
  \vspace{-6mm}
		\centering
    \centering
    \includegraphics[width=1.1\linewidth]{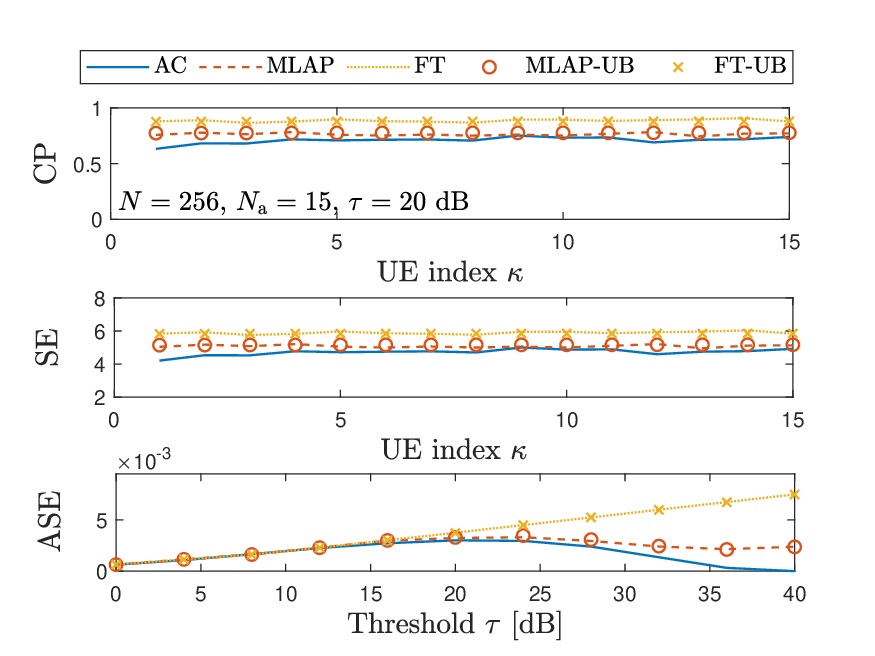}
    \vspace{-6mm}
    \caption{Approximate and actual values of SIR-related overall CP, overall SE, and ASE.}
    \label{fig:uncond_validate}
    \end{minipage}
\end{figure*}

 \begin{figure*}[t!]
 \vspace{-5mm}
\begin{minipage}{.325\textwidth}
    \centering
    \includegraphics[width=1.1\linewidth]{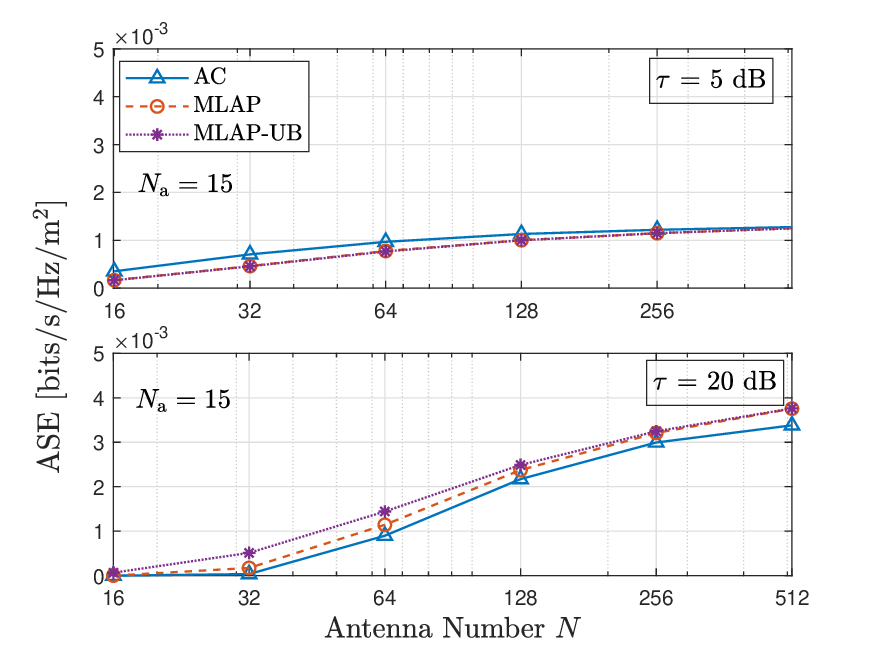}
    \vspace{-6mm}
    \caption{\textcolor{black}{Impact of the antenna number $N$ on the SIR-related ASE at different thresholds.}}
    \label{fig:ASE-Nt}
\end{minipage}
\hfill
\begin{minipage}{.325\textwidth}
    \centering
    \includegraphics[width=1.1\linewidth]{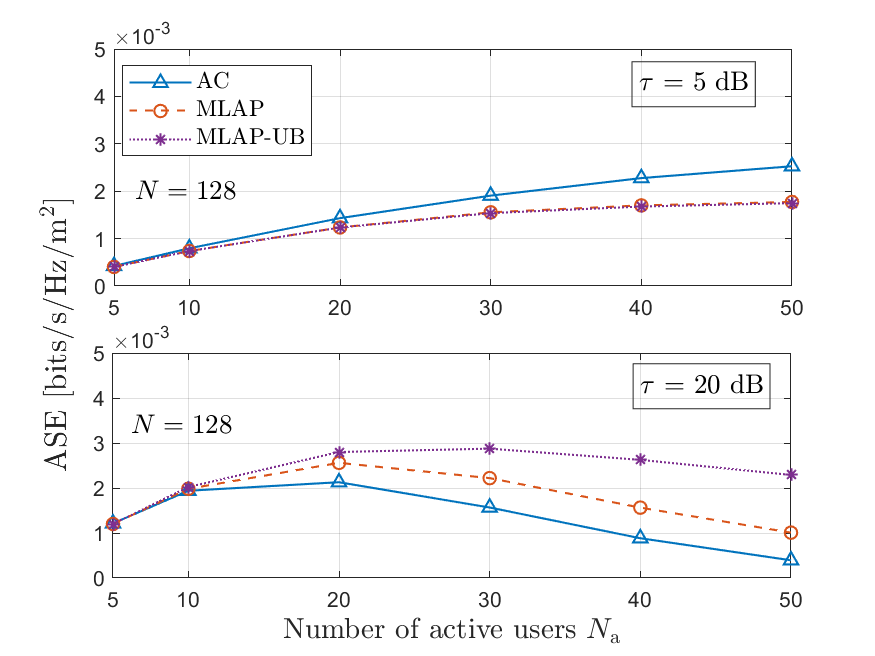}
    \vspace{-6mm}
    \caption{\textcolor{black}{Impact of the active-user number $N_{\rm a}$ on the SIR-related ASE at different thresholds.}}
    \label{fig:ASE-Na}
    \end{minipage}
\hfill
\begin{minipage}{.325\textwidth}
		\centering
    \centering
    \includegraphics[width=1.1\linewidth]{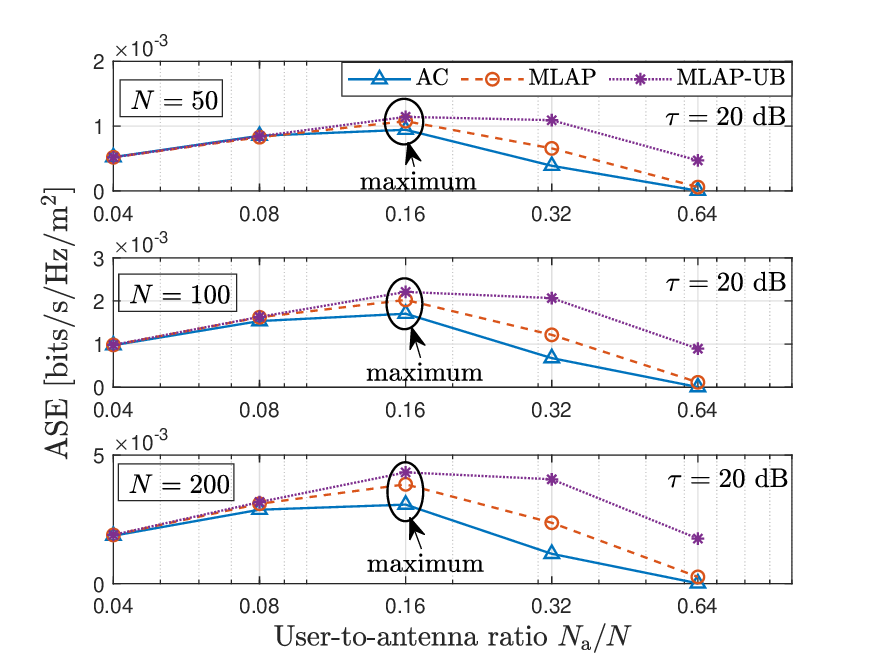}
    \vspace{-6mm}
    \caption{Impact of the user-to-antenna ratio $\frac{N_{\rm a}}{N}$ on the SIR-related ASE at $\tau=20~\rm dB$.}
    \label{fig:ASE-ratio}
    \end{minipage}
\end{figure*}

Fig.~\ref{fig:condCP} shows conditional \acp{CP} under different antenna patterns, given the $\kappa$-th active user located at $[\theta_\kappa,r_\kappa]^{\mathsf{T}}$.
For low thresholds ($\tau<5~\rm dB$), `AC' and `FT' nearly coincide, but their gap increases at the higher thresholds. 
This implies that the NF-MLAP with $M=1$ is only accurate in a limited SIR range since the flat-top approximation ignores sidelobe effects. The `MLAP' curve with $M=10$ provides a more accurate approximation than that with $M=1$. 
These results confirm the effectiveness of the NF-MLAP and suggest that a larger $M$ is needed to capture sidelobe effects at higher thresholds.
Moreover, we see that the `MLAP' and `MLAP-UB' (or `FT' and `FT-UB') curves are closely aligned, showing that the computationally-efficient upper bound in Theorem~\ref{theo:approxCP_upper} is tight.
Furthermore, when $\tau>35~\rm dB$ for the `MLAP' and `MLAP-UB' (or $\tau>15~\rm dB$ for the `FT' and `FT-UB'), the approximate CPs flatten and deviate notably from the actual CP. This observation is consistent with the discussion in Sec.~\ref{subsec:M}.

Fig.~\ref{fig:condCPsinr} compares the SIR and SINR-related conditional \acp{CP}. 
The noise power is set as $\sigma^2=-174 ~{\rm dBm/Hz}+ 10\log_{10} B + F~ {\rm dB}$, where the bandwidth is $B=200~\rm MHz$ and the noise figure is $F =10$~\cite{chen2024joint}.
For the actual pattern, SINR- and SIR-based CPs nearly coincide across a wide range of $\tau$. 
For example, the 3rd user at $(0^\circ,30~\rm m)$ exhibits overlapping `AC (SINR)' and `AC (SIR)' curves throughout $\tau\in[0,40]~\rm dB$, confirming the negligible impact of noise.
For users far from the BS, path loss increases and the received signal power decreases, making noise more noticeable at high thresholds. For example, the 12-th user at $(0^\circ,120~\rm m)$ shows slight divergence between `AC (SINR)' and `AC (SIR)' around $\tau\in[25,35]~\rm dB$.
Similar results can be observed from the NF-MLAP based approximations.
These observations confirm that the inter-user interference is the dominant factor for coverage performance across most thresholds, justifying the focus on SIR-based analysis in the subsequent evaluation.

\color{black}
Fig.~\ref{fig:value-M} presents the approximation accuracy of the NF-MLAP with different $M$. \textcolor{black}{The reference value $M^*(\tau)$ is obtained from Proposition~\ref{prop:M_scaling}, with $M^*(\tau)=2,3,7,12$ for $\tau=5, 20, 30,35~\rm dB$, respectively.}
We see that for a given threshold $\tau$, increasing $M$ up to $M^*(\tau)$ improves the approximation accuracy as more dominant sidelobes are included in the NF-MLAP. 
Beyond $M^*(\tau)$, further increasing $M$ offers negligible additional accuracy.
This indicates that a finite number of dominant lobes is sufficient for coverage analysis, and higher thresholds require larger $M$. \textcolor{black}{This behavior is consistent with Proposition~\ref{prop:M_scaling}.}
Moreover, Fig.~\ref{fig:value-M} shows that choosing $M \ge M^*(\tau)$ does not eliminate the approximation gap.
This is because the NF-MLAP is a discretized representation of a continuous antenna pattern in both angle and distance. 
This discretization introduces an inherent approximation error that cannot be eliminated simply by adding more sidelobes. 
\textcolor{black}{Overall, although $M^*(\tau)$ does not guarantee an exact match to the actual performance, it provides a sufficient condition for selecting $M$ to capture dominant interferers.} 

\color{black}

Fig.~\ref{fig:uncond_validate} provides the SIR-related overall \ac{CP}, overall \ac{SE}, and \ac{ASE}.
Comparing `MLAP' with `AC', we see that the approximate \ac{CP} and \ac{SE} align with the actual \ac{CP} and \ac{SE} at $\tau=20~\rm dB$; the approximate \ac{ASE} aligns with the actual \ac{ASE} for $\tau\in [0,30]~\rm dB$. This validates the approximation accuracy under the NF-MLAP with $M=10$.
Interestingly, the \ac{CP} and \ac{SE} are similar across active users, implying that each active user experiences a similar level of interference from the remaining $N_{\rm a}-1$ active users. 
Regarding the \ac{ASE}, it initially increases with $\tau$, reaches a peak at $20~\rm dB$, and then declines.
As shown in Sec.~\ref{subsec:metrics}, a higher $\tau$ allows each successful link (with SIR greater than $\tau$) to transmit more data, i.e., $\log_2(1+\tau)~\rm bits/s/Hz$, but fewer links can satisfy this stricter SIR requirement, leading to a reduced reliability, i.e., lower \ac{CP}. Here, $\tau=20~\rm dB$ represents an optimal operating point that balances communication reliability and rate.

\vspace{-2mm}
\color{black}
\subsection{Impact of System Parameters}\label{subsec:simu_para}

According to the characterization of the low- and high-threshold regimes in Sec.~\ref{subsec:low-high-tau}, in the following, we consider $\tau=5~\rm dB$ and
$\tau=20~\rm dB$ as representative relaxed and stringent SIR
requirements, respectively.

\subsubsection{Impact of Antenna Number}
Fig.~\ref{fig:ASE-Nt} illustrates the impact of the antenna number $N$ on the ASE.
First, we see that as $N$ increases, both the actual and approximate values of \ac{ASE} increase.
This increase can be attributed to the enhanced {spatial selectivity} and the resulting reduction of inter-user interference as $N$ increases. 
Moreover, we see that the ASE exhibits different behaviors under the two SIR thresholds.
\textcolor{black}{At $\tau=5~\rm dB$, the ASE increases with $N$ and gradually approaches a saturation trend, indicating that the system quickly approaches the interference-negligible regime under the relaxed SIR requirement.
In contrast, at $\tau=20~\rm dB$, the ASE continues to increase over the considered range of $N$ without reaching its saturation regime. 
As revealed in Sec.~\ref{subsec:ASEscale}, under a stringent SIR requirement, the interference penalty decays more slowly with $N$, and thus the system remains interference-limited over a wider range of antenna sizes.}

\subsubsection{Impact of Active-User Number} 

Fig.~\ref{fig:ASE-Na} shows the impact of the active-user number ($N_{\rm a}$) on the ASE under different SIR thresholds. 
\textcolor{black}{When $\tau=5~\rm dB$, all three curves (`AC', `MLAP', and `MLAP-UB') increase over the considered range of $N_{\rm a}$.  
This indicates that under the relaxed SIR requirement, the interference penalty is mild while the spatial reuse gain is dominant. 
For $\tau=20~\rm dB$, the ASE exhibits a clear unimodal behavior.}
Specifically, both the `AC' and `MLAP' curves peak at $N_{\rm a}=20$, while the `MLAP-UB' curve peaks slightly later at $N_{\rm a}=30$.
This indicates that, under more stringent SIR requirements, the interference penalty dominates the spatial reuse gain beyond the optimal $N_{\rm a}$, significantly limiting the number of simultaneously served users. 
These observations are consistent with the scaling laws in Sec.~\ref{subsec:ASEscale}, further confirming the effectiveness of the NF-MLAP.



\subsubsection{Optimal User-to-Antenna Ratio}  

Fig.~\ref{fig:ASE-ratio} further examines the ASE as a function of the ratio $\frac{N_{\rm a}}{N}$ for different antenna numbers at $\tau=20~\rm dB$.  
We see that for $N=50,100,200$, the actual ASE is consistently maximized at approximately the same ratio of $\frac{N_{\rm a}}{N}=0.16$, indicating that the optimal number of active users increases proportionally with $N$ when $\tau=20~\rm dB$.
This observation is consistent with the scaling law at Sec.~\ref{subsec:ASEscale}, i.e., the optimal $N_{\rm a}$ scales linearly with $N$ in the high-threshold regime.
Moreover, although the approximate ASE and its upper bound (i.e., the `MLAP' and `MLAP-UB' curves) deviate slightly from the actual ASE (i.e., the `AC' curve), they follow consistent trends and identify the same optimal ratio.
These results demonstrate that the proposed analytical framework offers computationally efficient yet accurate tools for system design.


 \vspace{-2mm}

\color{black}

\section{Conclusions}\label{sec:conclusion}

\color{black}
In this paper, we developed an NF-based SG framework to analyze spatial multiple access enabled by polar-domain beamfocusing, and proposed an approximate antenna pattern, namely, the NF-MLAP, for efficient performance evaluation.
Based on this framework, we established scaling laws that reveal how the system performance is governed by the interplay between spatial reuse and inter-user interference. In particular, both the benefit of increasing the antenna number and the admissible number of simultaneously served users depend on the SIR requirement, and there exists an optimal user-to-antenna ratio.
These findings provide useful insights for the design of NF multi-user systems. 

For future work, it is of interest to extend the proposed framework to account for beam misalignment due to imperfect or outdated location and channel information, \textcolor{black}{or due to beam selection from a predefined polar-domain codebook}, especially in dynamic scenarios with user mobility. 
Such an extension would enable the characterization of system performance under different levels of misalignment and provide insights for robust system design. 
Moreover, extending the proposed analytical framework to more advanced beamforming schemes, such as hybrid analog-digital beamforming, is also interesting.
Another direction is to consider alternative active-user distributions and corresponding user selection mechanisms, 
e.g., preferentially scheduling users that are well separated in the angle-distance domain based on their beamfocusing regions to mitigate inter-user interference.
\color{black}

 
\appendices

\vspace{-2mm}

\section{Proof of Theorem~\ref{them:cov}}\label{app:them:cov}

The conditional \ac{CP} in \eqref{eq:condCP} can be expressed as
\begin{align}\label{eq:condCP0}
{\rm CP}_{\kappa}(\tau |\theta_{\kappa},r_\kappa )=\mathbb{P}\{ {I_\kappa} < \frac{ 1 }{\tau} ~\big|~\theta_\kappa,r_\kappa\}= F_{I_{\kappa}}(\tau^{-1}),  
\end{align}
where $I_{\kappa} $ is interference defined in \eqref{eq:SIR_k}, and $F_{I_{\kappa}}(\cdot)$ is the \ac{CDF} of $I_{\kappa}$.
From Gil-Pelaez theorem~\cite{inversion}, we have
\begin{align} \label{eq:CDF_Ik}
	\begin{split}
		F_{I_{\kappa}}(w)
		&= \frac{1}{2}-\!\!  \int_{0}^{\infty} \!\!\frac{1}{\pi t} \, \operatorname{Im} \{ {\rm e}^{-{\rm j}  t w} \mathcal{L}_{\theta_{\kappa},r_\kappa}(-{\rm j} t) \} \mathrm{d}t,	
	\end{split}
\end{align}
where ${\rm Im}\{\cdot\}$ is the imaginary part of a complex number, and $\mathcal{L}_{\theta_{\kappa},r_\kappa}(s)=\mathbb{E}_{\Psi_{\rm a}|\theta_\kappa, r_\kappa}\left\{\exp{\left( -s  I_{\kappa}  \right) } \right\}$ is the Laplace transform of $I_{\kappa}$. 
Moreover, $I_{\kappa}$ in \eqref{eq:SIR_k} can be rewritten as
\begin{align}\label{eq:Ik}
    I_\kappa &=\!\!\!\!\!\!\!\!\!\!\!\sum_{\kappa',\mathbf{u}_{\kappa'}\in \Psi_{\rm a} \setminus \{\mathbf{u}_\kappa\}}  \!\!\!\!\!\!\!\!\!\!\mathcal{G} (\theta_\kappa,r_{\kappa};\theta_{\kappa'},r_{\kappa'})\overset{(a)}{=}\!\!\!\!\!\!\!\sum_{\mathbf{u}\in \Psi_{\rm a} \setminus \{\mathbf{u}_\kappa\}}  \!\!\!\!\!\!\mathcal{G} (\theta,r;\theta_\kappa,r_{\kappa})
   \nonumber \\&\overset{(b)}{=}  
   \sum_{\mathbf{u}\in \Psi^{\kappa}_{\rm in}  } \mathcal{G} (\theta,r;\theta_\kappa,r_{\kappa}) + 
   \sum_{\mathbf{u}\in \Psi^{\kappa}_{\rm out} } \mathcal{G} (\theta ,r ;\theta_\kappa,r_{\kappa} ), 
\end{align}
where (a) replaces $\mathbf{u}_{\kappa'}=[\theta_{\kappa'},r_{\kappa'}]^{\mathsf{T}}$ with $\mathbf{u}=[\theta,r]^{\mathsf{T}}$ for notational simplicity, (a) also exploits the symmetry of the inner product, and (b) follows from the partition in \eqref{eq:Psi_in_out}. 
Define $I_{v,\kappa}=\sum_{\mathbf{u}\in \Psi^{\kappa}_{v} } \mathcal{G} (\theta ,r ;\theta_\kappa,r_{\kappa} )$ as the interference caused by beams intended for active users located close ($v=\rm in$) or further ($v=\rm out$) than the $\kappa$-th active user to the BS. Then, the Laplace transform of $I_{\kappa}$ can be expressed as  
\begin{align}\label{eq:L_Ik}	\!\mathcal{L}_{\theta_{\kappa},r_\kappa}(s) \!=\! \mathbb{E}_{\Psi_{\rm a}|\theta_\kappa, r_\kappa} \!\left\{ \! {\rm e}^{ -s  (I_{{\rm in},\kappa} + I_{{\rm out},\kappa}) } \!  \right\}
\!=\!\!\!\!\!\!\!\!\!\prod_{v\in\{\rm in, out\}} \!\!\!\!\! \!\!\!L_{\theta_{\kappa},r_\kappa}^{v}(s),
\end{align}
where $L_{\theta_{\kappa},r_\kappa}^{v}(s) \triangleq\mathbb{E}_{\Psi_{v}^{\kappa}|\theta_\kappa, r_\kappa} \left\{{\rm e}^{ -s I_{{v},\kappa}}    \right\}$. 
Following the \ac{BPP} properties of $\Psi^{\kappa}_{\rm in}$ \cite[Appendixes B and C]{BPP1}, we obtain 
\begin{align}\label{eq:L_I_in}
	&L_{\theta_{\kappa},r_\kappa}^{\rm in}(s)  
        = \mathbb{E}_{\Psi_{\rm in}^{\kappa}|\theta_\kappa, r_\kappa} \left\{ {\rm e}^ {-s \sum_{\mathbf{u}\in \Psi^{\kappa}_{\rm in}} \mathcal{G} (\theta,r;\theta_\kappa,r_{\kappa}) }  \right\}
        \nonumber\\&=\mathbb{E}_{\Psi_{\rm in}^{\kappa}|\theta_\kappa, r_\kappa}\bigg \{ \prod_{\mathbf{u}\in \Psi^{\kappa}_{\rm in}} {\rm e}^{ -s \mathcal{G} (\theta,r;\theta_\kappa,r_{\kappa}) }   \bigg\} 
		\nonumber\\&=  \bigg( \!  \int_{-\frac{\pi}{N_{\rm s}}}^{\frac{\pi}{N_{\rm s}}} \! \int_{0}^{r_{\kappa}} {\rm e}^{ -s \mathcal{G} (\theta,r;\theta_\kappa,r_{\kappa})} f_{R_{\rm in}|r_{\kappa}}(r) \frac{N_{\rm s}}{2\pi}\mathrm{d} r \mathrm{d} \theta\bigg)^{\kappa-1} \!\!\!\!\!.
\end{align}
The derivation of $L_{\theta_{\kappa},r_\kappa}^{\rm out}(s)$ follows similarly. 
Finally, substituting $L_{\theta_{\kappa},r_\kappa}^{\rm in}(s)$ and $L_{\theta_{\kappa},r_\kappa}^{\rm out}(s)$ into \eqref{eq:L_Ik} and \eqref{eq:CDF_Ik} completes the proof of Theorem~\ref{them:cov}.

\vspace{-2mm}

\section{Proof of Theorem~\ref{theo:approxCP_upper}}\label{app:theo:approxCP_upper}
Substituting \eqref{eq:Ik} into \eqref{eq:condCP0} and replacing the actual antenna pattern with the NF-MLAP, we obtain
\begin{align}\label{eq:CPcond_upper1}
		&\widetilde{\rm CP}_{\kappa}(\tau |\theta_{\kappa},r_\kappa )
         = \mathbb{P}\bigg\{  \underbrace{ \sum_{\mathbf{u}\in \Psi_{\rm a} \setminus \{\mathbf{u}_\kappa\}}  \!\!\!\!\!\! \widetilde{\mathcal{G}}_{M} (\theta,r;\theta_\kappa,r_{\kappa}) < \frac{ 1 }{\tau} }
_{\text{event }\mathcal{A}}~\bigg|~\theta_\kappa,r_\kappa\bigg\} 
\nonumber\\& \overset{(a)}{\le}   
 \mathbb{P}\bigg\{ \underbrace{ \bigcap_{\mathbf{u}\in \Psi_{\rm a} \setminus \{\mathbf{u}_\kappa\}} \bigg\{ \widetilde{\mathcal{G}}_{M}(\theta,r;\theta_\kappa,r_{\kappa}) < \frac{ 1 }{\tau} \bigg\} }
_{\text{event }\mathcal{B}}~\bigg|~\theta_\kappa,r_\kappa\bigg\}
		\nonumber\\&  \triangleq  \widehat{\rm CP}_{\kappa}(\tau |\theta_{\kappa},r_\kappa )  ,
\end{align}
where (a) is from $\mathbb{P}(\mathcal{A})\le \mathbb{P}(\mathcal{B})$ since $\mathcal{A} \subset \mathcal{B}$, i.e., if the sum of non-negative random variables $\widetilde{\mathcal{G}}_{M} (\theta,r;\theta_\kappa,r_{\kappa})$ is less than 
$\frac{1}{\tau}$, then each individual term must also be less than $\frac{1}{\tau}$.
The upper bound in \eqref{eq:CPcond_upper1} can be further expressed as
\begin{align}\label{eq:CPcond_upper2}
		&\widehat{\rm CP}_{\kappa}(\tau |\theta_{\kappa},r_\kappa )
		\nonumber\\& \overset{(a)}{=} \prod_{\mathbf{u} \in \Psi_{\rm a} \setminus \{\mathbf{u}_\kappa\}}  
 \mathbb{P}\left\{ \widetilde{\mathcal{G}}_{M}(\theta ,r ;\theta_\kappa,r_{\kappa}) < \frac{ 1 }{\tau} ~\bigg|~\theta_\kappa,r_\kappa\right\} 
\nonumber \\& \overset{(b)}{=}  
 \prod_{v\in\{\rm in, out\} } \prod_{\mathbf{u}\in \Psi^\kappa_{v} }  
 \mathbb{P}\left\{ \widetilde{\mathcal{G}}_{M}(\theta,r;\theta_\kappa,r_{\kappa}) < \frac{ 1 }{\tau} ~\bigg|~\theta_\kappa,r_\kappa\right\} 
\nonumber \\& \overset{(c)}{=}  
 \bigg(\underbrace{\mathbb{P}\left\{ \widetilde{\mathcal{G}}_{M}(\theta,r;\theta_\kappa,r_{\kappa}) < \frac{ 1 }{\tau} ~\bigg|~\theta_\kappa,r_\kappa,r<r_{\kappa}\right\} }_{P_{\rm in} (\tau)}\bigg)^{\kappa-1} \!\!\!\!\!\!  \times 
\nonumber \\&\quad ~\bigg(\underbrace{\mathbb{P}\left\{ \widetilde{\mathcal{G}}_{M}(\theta,r;\theta_\kappa,r_{\kappa}) < \frac{ 1 }{\tau} ~\bigg|~\theta_\kappa,r_\kappa,r>r_{\kappa}\right\} }_{P_{\rm out} (\tau)} \bigg)^{N_{\rm a}-\kappa}\!\!\!\!\!\!\!\!\!\!,    
\end{align}
where (a) is from the conditional independence of $\mathbf{u}=[\theta,r]^{\mathsf{T}}\in \Psi_{\rm a} \setminus \{\mathbf{u}_\kappa\}$ given $\mathbf{u}_\kappa=[\theta_\kappa,r_\kappa]^{\mathsf{T}}$, and (b) is from the partition in \eqref{eq:Psi_in_out}, and  
(c) is from the conditions $r<r_{\kappa}$ for $\mathbf{u}\in \Psi_{\rm in}^{\kappa}$ and $r>r_{\kappa}$ for $\mathbf{u}\in \Psi_{\rm out}^{\kappa}$.
Based on Lemma~\ref{lemma:prob_Gain},  $P_v(\tau)$, for $v\in\{\rm in, out\}$, is given by
\begin{align}\label{eq:P_tau}
    P_{v} (\tau) &= \sum_{i=0}^{M+1}\mathbbm{1}\left(g_i<\frac{1}{\tau}\right) p_i^{v},
\end{align}
where $\mathbbm{1}(\cdot)$ is the indicator function, and $p_i^{\rm in }$ (or $p_i^{\rm out }$) is the probability that $\widetilde{\mathcal{G}}_{M}(\theta,r;\theta_\kappa,r_{\kappa})=g_i$ for $r<r_\kappa$ (or $r>r_\kappa$), as derived in Lemma~\ref{lemma:prob_Gain}. 
Finally, substituting \eqref{eq:P_tau} into \eqref{eq:CPcond_upper2} completes the proof of Theorem~\ref{theo:approxCP_upper}.

 
\section{Proof of Corollary~\ref{cor:upper_eq}} \label{app:cor:upper_eq}
From \eqref{eq:CPcond_upper1} and \eqref{eq:NF_MLAP}, the possible non-zero values of $ \widetilde{\mathcal{G}}_{M} (\cdot)$ are $\{g_0,g_1,..., g_M\}$.
Let $\frac{1}{\tau^*}=\min_{i\in \{0,1,...,M\}} {g_i}$ equals the smallest non-zero gain. When $\tau=\tau^*$ we have 
\begin{align}
    \left\{\widetilde{\mathcal{G}}_{M} (\theta,r;\theta_\kappa,r_{\kappa})<\frac{1}{\tau^*}\right\} \Leftrightarrow  \{\widetilde{\mathcal{G}}_{M} (\theta,r;\theta_\kappa,r_{\kappa})=0\}.
\end{align} 
This indicates that the events $\mathcal{A}$ and $\mathcal{B}$ defined in   \eqref{eq:CPcond_upper1} are equivalent when $\tau=\tau^*$.
Furthermore, for $\tau>\tau^*$, any non-zero gain $g_i$ still satisfies $g_i \ge \frac{1}{\tau^*} > \frac{1}{\tau}$.
Thus, both events $\mathcal{A}$ and $\mathcal{B}$ remain equivalent for all $\tau> \tau^*$.
Hence, the equality in \eqref{eq:CPcond_upper1} holds 
when $\tau\ge\tau^*$, i.e.,  
\begin{align}\label{eq:approx_CP_const0}
& \widetilde{\rm CP}_{\kappa}(\tau |\theta_{\kappa},r_\kappa )= 
     \widehat{\rm CP}_{\kappa}(\tau |\theta_{\kappa},r_\kappa  )\nonumber\\& \overset{(a)}{=} \left(\sum_{i=0}^{M+1}\mathbbm{1}\left(g_i<\frac{1}{\tau}\right) p_i^{\rm in }\right)^{\kappa-1} \!\!\!\left(\sum_{i=0}^{M+1}\mathbbm{1}\left(g_i<\frac{1}{\tau}\right) p_i^{\rm out }\right)^{N_{\rm a}-\kappa} 
    \nonumber\\&  \overset{(b)}{=} \left( p_{M+1}^{\rm in } \right)^{\kappa-1} \left(  p_{M+1}^{\rm out } \right)^{N_{\rm a}-\kappa}, \text{ for } \tau \ge \tau^*,
\end{align}
where (a) is based on \eqref{eq:CPcond_upper2}-\eqref{eq:P_tau}, and (b) is from $g_i\ge\frac{1}{\tau^*}\ge\frac{1}{\tau}$ for $i=0,...,{M}$ and from $\frac{1}{\tau}>g_{M+1}=0$. 
This completes the proof of Corollary~\ref{cor:upper_eq}.

	\bibliographystyle{IEEEtran}
	\bibliography{reference}
\end{document}